\documentclass[fleqn,usenatbib]{mnras}

\usepackage{newtxtext,newtxmath}

\usepackage[T1]{fontenc}

\DeclareRobustCommand{\VAN}[3]{#2}
\let\VANthebibliography\thebibliography
\def\thebibliography{\DeclareRobustCommand{\VAN}[3]{##3}\VANthebibliography}

\usepackage{graphicx}	
\usepackage{amsmath}	
\usepackage{subfig}
\usepackage{array,multirow,graphicx}

\def\arcsec{\hbox{$^{\prime\prime}$}}
\def\utw{\smash{\rlap{\lower5pt\hbox{$\sim$}}}}
\def\udtw{\smash{\rlap{\lower6pt\hbox{$\approx$}}}}

\def\fdg{\hbox{$.\!\!^\circ$}}

\newcommand{\fermi}{\textit{Fermi}-LAT}

\newcommand{\angstrom}{\textup{\AA}}

\newcommand{\Rblob}{$R_{\rm blob,\,BLR}$}
\newcommand{\FluxFermiLC}{$\overline{\phi}_{\rm HE, 12}$}

\title[Constraints of VHE $\gamma$-ray emission of FSRQ with MAGIC]{Constraints on VHE gamma-ray emission of Flat Spectrum Radio Quasars with the MAGIC telescopes}

\author[S.~Abe~et.~al.]{\parbox{\textwidth}{\Large{
S.~Abe$^{1}$,
J.~Abhir$^{2}$,
A.~Abhishek$^{3}$,
V.~A.~Acciari$^{4}$,
A.~Aguasca-Cabot$^{5}$,
I.~Agudo$^{6}$,
T.~Aniello$^{7}$,
S.~Ansoldi$^{8,42}$,
L.~A.~Antonelli$^{7}$,
A.~Arbet Engels$^{9}$
% ...
N.~Żywucka$^{15}$
\vspace{0.4cm}\\
\textit{(Affiliations can be found after the references)}
\vspace{4.0cm}\\
}}
}

\newpage

\date{Accepted XXX. Received YYY; in original form ZZZ}

\pubyear{2020}

\begin{document}
\label{firstpage}
\pagerange{\pageref{firstpage}--\pageref{lastpage}}
\maketitle
\begin{abstract}
Flat spectrum radio quasars (FSRQs) constitute a class of jetted active galaxies characterized by a very luminous accretion disk, prominent and rapidly moving line-emitting cloud structures (Broad Line Region, BLR), and a surrounding dense dust structure known as dusty torus. The intense radiation field of the accretion disk strongly determines the observational properties of FSRQs. 
While hundreds of such sources have been detected at GeV energies, only a handful of them exhibit emission in the very-high-energy (VHE, E $\gtrsim 100$ GeV) range.

This study presents the results and interpretation derived from a cumulative observation period of 174 hours dedicated to nine FSRQs conducted with the MAGIC telescopes from 2008 to 2020. Our findings indicate no statistically significant ($\geq$ 5 $\sigma$) signal for any of the studied sources, resulting in upper limits on the emission within the VHE energy range.

In two of the sources, we derived quite stringent constraints on the $\gamma$-ray emission in the form of upper limits.
Our analysis focuses on modeling the VHE emission of these two sources in search for hints of absorption signatures within the broad line region (BLR) radiation field. For these particular sources, constraints on the distance between the emission region and the central black hole are derived using a phenomenological model. Subsequently, these constraints are tested using a framework based on a leptonic model.
 
\end{abstract}




\section{Introduction}
Active galactic nuclei (AGN) are one of the most luminous class of objects in the Universe. Jetted AGN whose jet is pointing towards or making a small angle with respect to the observer are called blazars. 
Based on their optical properties, blazars are divided into two groups: BL Lacertae (BL Lac) objects and flat spectrum radio quasars (FSRQs). Blazars with strong optical emission lines having an equivalent width (EW) of EW > 5 {\AA} are classified as FSRQs, while those with weak or absent lines are referred to as BL Lacs \citep{1995PASP..107..803U}. The distinction between FSRQ and BL Lac is unclear in some cases due to the lack of good quality measured optical spectra or changing EW of the lines during high states. 
Broad emission lines identifiable in the spectra of some AGNs are produced in the broad line region (BLR), which is considered photoionized by thermal radiation from the accretion disk \citep{2007A&A...464..871R}.

Two humps characterize the broadband spectral energy distribution (SED) of blazars. The lower energy hump in the spectrum is caused by synchrotron emission produced by relativistic electrons within the jet. On the other hand, the higher energy component, typically peaking at GeV-TeV, is often interpreted as inverse Compton (IC) scattering. This scattering occurs between the relativistic electrons and either the synchrotron photons themselves (synchrotron self Compton, SSC) or external photons outside the jet (external Compton, EC). The SSC scenario is commonly used to explain the higher energy peak in BL Lac objects, while the EC scenario is favored for FSRQs \citep{2014ApJ...790...45P}. The external photons are produced in the dusty torus (DT), in the BLR, or could come directly from the accretion disk \citep{2002ApJ...575..667D, 2007ASPC..373..169B, 2018ApJ...863...98P, 2019ApJ...874...47V}. The nature of the dominating external radiation field depends on the location of the emission zone with respect to the black hole.
When the emission region is located within the BLR, a sharp cut-off in the $\gamma$-ray spectrum is foreseen to occur due to the strong attenuation of the high-energy (HE, order of a few GeV) and very-high-energy (VHE, E $\gtrsim 100$ GeV) $\gamma$-rays through their interaction with the optical photons in the $\gamma$-$\gamma$ pair production process. The opacity for HE photons could be very large, averting their escape from the emission region; therefore, if we can attribute the cause of the break/cut-off to the absorption, we can put a constraint on the location of the $\gamma$-ray emitting zone \citep{2020A&A...635A..25S}. 

Thus, the $\gamma$-ray emission from FSRQs can be affected by internal absorption in the dense UV-optical photons of the BLR (\citealp{2006ApJ...653.1089L}, \citealp{2018MNRAS.477.4749C}). Detection of $\gamma$-rays from blazars by imaging atmospheric Cherenkov telescopes (IACTs) is also made difficult due to the interaction between VHE photons and lower energy extragalactic background light (EBL) photons \citep{2010ApJ...712..238F, 2011MNRAS.410.2556D, 2021MNRAS.507.5144S, 2022ApJ...941...33F}, which also leads to electron-position pair creation. This process strongly attenuates VHE photons with energies above a characteristic energy, which depends on the redshift of the source, typically about 30 GeV for sources at redshift z $<$ 1. The attenuation results in a softer spectrum observed on Earth than the intrinsic spectrum of extragalactic sources \citep{1962JPSJS..17C.175G}, making the detection of distant sources challenging \citep[see e.g.,][]{2014MNRAS.440..530A}. 

The fourth catalog of AGN detected by \textit{Fermi}-LAT (4LAC-DR3,  \citealp{2022ApJS...263..24A}) contains 2896 sources, among which 640 are FSRQs, and 1261 are BL Lacs. Based on the results from 4LAC, on average, FSRQs demonstrate softer spectra and stronger variability in the $\gamma$-ray energy range than BL Lacs, confirming previous results  \citep{2013SAAS...40..225D}. However, it is important to note that the stronger variability might be an observational bias due to the tendency of FSRQs to be brighter.
The catalogs of hard \textit{Fermi}-LAT sources (2FHL \citealp{2016ApJS..222....5A} and 3FHL \citealp{2017ApJS..232...18A}) report detections of FSRQs at different energy thresholds; the number of FSRQs detected above 10 GeV in the 3FHL (integration time $\sim$ 84 months) is 172, and the number of FSRQs detected above 50 GeV in the 2FHL (integration time $\sim$ 80 months) is only 10.

FSRQs are generally located at higher cosmological distances than BL Lacs \citep{2020ApJS..247...33A}, implying a strong absorption of $\gamma$-rays by the EBL (see e.g. PKS 1441+25, \citealp{2015ApJ...815L..23A}). The most distant (with redshift $z \sim 1$) AGNs ever detected in the VHE energy range are the FSRQ PKS 1441+25 at $z=0.940$ \citep{2015ApJ...815L..23A}, the FSRQ PKS 0346-27 at $z=0.991$ \citep{2018ATel11251....1A, 2021ATel15020....1W} and the gravitationally lensed blazar QSO B0218+357 \citep{2016A&A...595A..98A} at $z=0.954$. At the end of 2023, LST-1 announced detection VHE emission from the FSRQ OP 313 \citep{2023ATel16381....1C} with redshift $z = 0.997$ \citep{2010AJ....139.2360S}. 

FSRQs are highly variable in the VHE band (see, e.g., \citealp{2019ApJ...877...39M}). The VHE $\gamma$-ray flux has been observed to fluctuate even by two orders of magnitude \citep{2019Galax...7...41Z}. Due to this, the most successful approach for studying the FSRQs in the VHE $\gamma$-rays is to follow up alerts of enhanced activity at lower energies. However, it should be kept in mind that this strategy is only effective if the flare is of long duration and not on hour-timescale. Additionally, continuous monitoring is necessary for this approach to be fruitful.

The first catalog reporting upper limits (ULs) from FSRQs in the TeV energy range was carried out by the Whipple Collaboration \citep{2004ApJ...613..710F}. A catalog of ULs for AGNs, including FSRQs sources, was also published by the High Energy Stereoscopic System Collaboration \citep[H. E. S. S., ][]{2008A&A...478..387A,2014A&A...564A...9H} and the Very Energetic Radiation Imaging Telescope Array System \citep[VERITAS, ][]{2016AJ....151..142A}. Nowadays, the VERITAS collaboration performs a systematic and unbiased search for the TeV emission from a set of FSRQs \citep{2021arXiv210806099P}.

 
 The current list of detected blazars in the VHE band, available in \textit{TeVCat}\footnote{\url{http://tevcat.uchicago.edu/}} \citep{2008ICRC....3.1341W}, consist of 70 BL Lacs and only 9 FSRQs.   



Observations of FSRQs in the VHE band may provide information about their nature and radiation processes. The longstanding question pertains to the location of the $\gamma$-ray emitting region within FSRQs. Recent evidence suggests that this region is likely located beyond the BLR, at least during the epoch of VHE $\gamma$-ray emission. Such evidence stems both from the absence of absorption in the \textit{Fermi}-LAT observations of FSRQs - where 2/3 of the selected FSRQs in the study of \cite{2018MNRAS.477.4749C} displayed no signs of absorption within the $>$100 GeV range, as well as from the detection of VHE $\gamma$-ray emissions from FSRQs. FSRQs experience variability in VHE $\gamma$-rays with timescales as low as tens of minutes (see \citealp{2011ApJ...730L...8A, 2017ICRC...35..655Z}). It has been long argued that such variability would more naturally occur closer to the black hole; however, as it has become evident that jets have substructures and that the emission region does not have to fill the full jet diameter (see, e.g., \citealt{2019NewAR..8701541H} for a recent review), this line of argumentation has become less popular. Nowadays, the understanding of VHE emission of FSRQs still needs to be completed. 

In this paper, we present the VHE $\gamma$-ray observations, data analysis, and results of nine FSRQs: 
TXS 0025+197, B2 0234+28, AO 0235+16, 4C+55.17, OP\,313, CTA 102, B2 2234+28A, TXS 2241+406, 3C 454.3.
The data used in this study were gathered by the Major Atmospheric Gamma-ray Imaging Cherenkov telescopes (MAGIC) together with the optical data from the Kungliga Vetenskapsakademien (KVA) along with X-ray, UV, and optical data from \textit{Swift}-XRT and \textit{Swift}-UVOT, respectively. We also performed dedicated \textit{Fermi}-LAT analysis contemporaneous to the MAGIC observations.
Additionally, we used $\gamma$-ray data collected by \textit{Fermi}-LAT over 12 years to compare these observations with an average state of the sources studied in this paper. We present here a MAGIC catalog of ULs on the $\gamma$-ray emission of these sources. Next, we construct a theoretical model using the \textit{Fermi}-LAT data and the MAGIC ULs exploiting the absorption in BLR and, finally, derive a broadband emission model based on the EC scenario.\\
The paper is organized in the following way: a description of instruments and data analysis method is included in Section~\ref{sec:inst}, notes on the individual sources are reported in Section~\ref{sec:srcs}, $\gamma$-ray emission results are discussed in Section~\ref{sec:results}, constraints on the distance between the emission region and the central black hole are described in Section~\ref{sec:model}, along with broadband modeling. Finally, the results are summarized in Section~\ref{sec:summary}. In the Appendix, we provide the light curves (LCs) for seven out of nine sources and SEDs for all the studied sources.  

\section{Instruments, Observations, and Data Analyses}\label{sec:inst}
We investigate the broadband emission of the sources by using data from the following instruments: MAGIC (VHE $\gamma$-rays), \fermi\ (GeV $\gamma$-rays), KVA (optical band) as well as, for selected sources, \textit{Swift}-UVOT (UV and optical bands) and \textit{Swift}-XRT (X-rays).

\subsection{MAGIC}
\label{sec:magic}
MAGIC is a stereoscopic system consisting of two 17-m diameter IACTs located at Observatorio del Roque de los Muchachos on the Canary Island of La Palma \citep{2016APh....72...61A}. The MAGIC telescopes are able to reach a low energy threshold of 50 GeV at low zenith angles \citep{2016APh....72...76A}.
Due to such low energy thresholds, they are well-suited for studies of high-redshift blazars. Some of the data used in this study were taken with a standard trigger, and the rest were taken with a special low-energy analog trigger called Sum-Trigger-II, designed to improve the performance of the telescopes reaching an even lower energy threshold of $\sim$ 15 GeV \citep{2021ITNS...68.1473D}. 
Sum-Trigger-II in a stereoscopic system allows the combination of the low energy trigger threshold along with better background rejection compared to standard stereo trigger. 
It also requires a special analysis procedure to get a larger effective area at lower energies. At higher energies, SUM-Trigger-II has a smaller trigger region. Therefore, it is used only with selected low-energy sources, particularly those located at high redshifts. The observations were carried out in wobble mode \citep{1994APh.....2..137F} with a 0.4$^{\circ}$ offset and four symmetric positions distributed around the camera center, improving the statistical accuracy of background estimation.
The data selection was based on the atmospheric transmission measured mainly with light detection and ranging (LIDAR) system \citep{2023A&A...673A...2S} and rates of background events. The data were analyzed using the MAGIC Analysis Reconstruction Software (MARS) framework \citep{2013ICRC...33.2937Z,  2016APh....72...76A}. 
For each source, all of the data available, therefore also from the different observation periods and consequently flux states,  were combined together to obtain a constraint on its VHE emission.
The ULs were calculated using the method presented in \cite{2005NIMPA.551..493R}, with a 95\% Confidence Level (C.L.). This approach assumes a systematic Gaussian uncertainty in the detector efficiency, i.e., the effective area, with a $\sigma$ of 30\%.

\subsection{\fermi}
\label{sec:fermi}
The Large Area Telescope onboard the \textit{Fermi} satellite is a gamma-ray instrument that detects photons by conversion into electron-positron pairs and has an operational energy range from 20\,MeV to more than 300\,GeV. It comprises a high-resolution converter tracker (for direction measurement of the incident $\gamma$-rays), a CsI(Tl) crystal calorimeter (for energy measurement), and an anti-coincidence shield detector to identify and veto the background of charged particles \citep{2009ApJ...697.1071A}.
We performed a dedicated analysis of the \textit{Fermi}-LAT data for each of the nine MAGIC-observed FSRQs using 12 years of LAT observations taken between 4th August 2008 and 4th August 2020. Additionally, as we will describe later, for each of the sources we analyzed the LAT data centered on the times of the MAGIC observations (see Appendix A for the exact times). A similar analysis technique has also been applied in \citet{2021MNRAS.507.4564P}.

We selected P8R3 SOURCE class events \citep{2018arXiv181011394B}, in the energy range between 100\,MeV and 1\,TeV, in a region of interest (ROI) of 15$^{\circ}$ radius centered on the position of each selected source. 
The value of the low energy threshold is motivated by the large uncertainties in the arrival directions of the photons below 100 MeV \citep{2018A&A...618A..22P}, leading to possible confusion between point-like sources and the Galactic diffuse component.

The analysis (which consists of model optimization, localization, spectrum and variability study) was performed with \texttt{Fermipy}\footnote{a Python package that facilitates analysis of LAT data with the \textit{Fermi} Science Tools \url{http://fermipy.readthedocs.io/en/latest/}} (version 1.0.1) \citep{2017ICRC...35..824W}, and the \textit{Fermi} Science Tools (version 11-07-00). The count maps were created with a pixel size of $0.1^{\circ}$. 
All $\gamma$-rays with zenith angle larger than 95$^{\circ}$ were excluded to limit the contamination from secondary $\gamma$-rays from the Earth’s limb \citep{2009PhRvD..80l2004A}. 
We made a harder cut at low energies by reducing the maximum zenith angle and by excluding event types with the worst point spread functions\footnote{A measure of the quality of the direction reconstruction is used to assign events to four quartiles. The $\gamma$-
rays in Pass 8 data can be separated into 4 PSF event types: 0, 1, 2, 3, where PSF0 has the largest point spread function, and PSF3 has the best.} (PSF).
Namely, for energies below 300 MeV, we excluded events with a zenith angle larger than 85$^{\circ}$, as well as photons from the PSF0 event type, while above 300 MeV we used all event types.
The P8R3\_SOURCE\_V3 instrument response functions (IRFs) are used.
The model used to describe the sky includes all point-like and extended LAT sources, located at a distance $<20^{\circ}$ from each FSRQ position, listed in the Fourth \textit{Fermi}-LAT Source Catalog  \citep[4FGL-DR2,][]{2020ApJS..247...33A}, as well as the Galactic diffuse and isotropic emission.
For these two latter contributions, we made use of the same templates\footnote{\url{https://fermi.gsfc.nasa.gov/ssc/data/access/lat/BackgroundModels.html} diffuse model: gll\_iem\_v07.fits and isotropic: iso\_P8R3\_SOURCE\_V3\_v1.txt} adopted to compile the 4FGL-DR2.
For the analysis, we first optimized the model for the ROI, then we searched for the possible presence of new sources, and finally, we re-localized the source.

We investigated the possible presence of additional faint sources, not in 4FGL-DR2, by generating test statistic\footnote{The test statistic is the logarithmic ratio of the likelihood of a source being at a given position in a grid to the likelihood of the model without the source, 
TS=2log($\mathrm{likelihood_{src}}/\mathrm{likelihood_{null}}$) 
\citep{1996ApJ...461..396M}} (TS) maps. No new bright sources were detected in the vicinity of our targets.
We left free to vary the diffuse background and the spectral parameters of the sources within 5$^{\circ}$ of our targets.
For the sources at a distance between 5$^{\circ}$ and 10$^{\circ}$, only the normalization was fitted, while we fixed the parameters of all the sources within the ROI at larger angular distances from our targets. The spectral fit was performed over the energy range from 100\,MeV to 1\,TeV.
To study the variability of the $\gamma$-ray emission of each FSRQ we divided the \textit{Fermi}-LAT data into time intervals of one week. For the light curve analysis, we fixed the photon index to the value obtained for 12 years and left only the normalization free to vary. The 95\% upper limit is reported for each time interval with TS $<10$. 

In addition to the study of the whole 12 years of \textit{Fermi}-LAT data, we performed a stacking analysis selecting and folding together all the photons observed in the considered periods, chosen to be simultaneous (around) to the MAGIC observations (for the exact time considered for the LAT analysis see Table. \ref{tab:magic_fermi_obs} and Fig. \ref{fig:plot_LC_Fermi_12year} in the Appendix). 
In particular, for each source we selected all the \textit{Fermi}-LAT data in daily (24 hours) time intervals centered on each MAGIC observation. Similarly to the procedure used for the MAGIC observations (see Sect. \ref{sec:magic}), where no analysis were carried on individual observations/nights but were all stacked together, we stacked all the data from the daily-intervals around each MAGIC observation. The one-day interval was chosen so that there would be enough LAT exposure for source detection and spectral reconstruction. For this study we adopted the same data selections and analysis procedure applied to the 12-years analysis (see above for the analysis description).
\subsection{KVA}
In the optical band, the sources were monitored by a 35 cm Celestron telescope attached to the KVA telescope as a part of the Tuorla Blazar Monitoring Program \citep{2008AIPC.1085..705T}. The monitoring program started in 2002 and was originally focused on TeV candidate BL~Lac objects from \cite{2002A&A...384...56C}, but the monitoring sample has been gradually increasing throughout the years. The monitoring observations are typically performed twice a week. However, as most of the sources in this paper are not part of the main sample, the cadence of the observations is poorer in some cases. The observations were performed using the Cousins R-filter.

The data were analyzed using the semi-automatic pipeline for differential photometry, developed for this purpose \citep{2018A&A...620A.185N}. For AO~0235+16 we used the comparison and control star magnitudes from \cite{2007A&A...464..871R} while for other sources we calibrated the
comparison stars using the observations of sources with known
comparison star magnitudes from the same night. 

As the sources have high redshift and bright optical nuclei, the contribution of the host galaxy to optical flux is negligible and we did not corrected for it. 
The measured magnitudes were corrected for galactic extinction using the galactic extinction model of \cite{2011ApJ...737..103S}. 

\subsection{\textit{Swift}-UVOT and \textit{Swift}-XRT} \label{data_analyssis:swift}

The \textit{Neil Gehrels Swift Observatory} \citep{2004ApJ...611.1005G} is a space satellite launched in 2004 by the National Aeronautics and Space Administration (NASA). It is equipped with three telescopes, namely the Ultraviolet and Optical Telescope \citep[UVOT; ][]{Roming05}, the Burst Alert Telescope \citep[BAT; ][]{Barthelmy05}, and the X-ray Telescope \citep[XRT; ][]{Burrows05}, initially built to monitor the $\gamma$-ray bursts and their afterglow phase, and then eventually developed into a versatile tool for collecting data in optical, UV, and X-rays from any source. Due to the presence of multiple instruments and rapid response to alerts, the \textit{Swift} observatory is ideal for gathering simultaneous data in multiwavelength (MWL) campaigns.

In this work, we performed the spectral analysis contemporaneous to the MAGIC observations and derived the long-term LCs for two of the sources, namely CTA 102 and B2 2234+28A. Both sources were monitored in the U (345 nm), B (439 nm), and V (544 nm) optical bands, and in the UVW2 (188 nm), UVM2 (217 nm), and UVW1 (251 nm) UV regime, as well as in X-ray energies between 0.2 and 10 keV. The comprehensive analysis was performed in the \textsc{ISIS} environment with \textsc{HEASOFT} version 6.30. 

The \textit{Swift}-UVOT instrumental magnitudes were calculated within a circular region centered at the source coordinates with a radius of 5\arcsec, using the \textsc{uvotsource} task. For the background determination, an annulus region centered at the same coordinates with an inner radius of 26\arcsec and an outer radius of 40\arcsec was used. The choice was made to prevent signal contamination from other sources in the closest vicinity of the studied blazars. Finally, we derived the fluxes taking into account the Galactic extinction $A_V$ correction based on the hydrogen absorption column density $N_H$ in the direction of the object and using the color excess E(B-V), calculated as $E(B-V) = N_H/(1.79 \times 10^{21} A_V)$ \citep{Predehl95}. 

CTA~102 (338$^\circ$15, 11$^\circ$73) was visible in three observation IDs, i.e., 00033509098, 00033509106, and 00033509110 for both UVOT and XRT. The LCs are generated with a longer time span from MJD 57624.9 to 57753.1 (2016-08-24 to 2016-12-31) based on 58 observation IDs between 00033509018 and 00088026001. The Galactic extinction was corrected with $N_H = 6.64 \times 10^{20}$cm$^{-2}$ \citep{Evans14}. 

Just one observation with 00038408004 ID is available for B2~2234+28A (339\fdg09, 28\fdg48) in the MAGIC time windows. While only one observation was considered for creating the broadband source spectrum, for the variability investigation, the LC was calculated between MJD 58641.1 and MJD 58668.3 (2019-06-07 to 2019-07-04) from two observations, namely 00038408002 and 00038408004. The Galactic extinction was corrected with $N_H = 6.15 \times 10^{20}$cm$^{-2}$ \citep{Evans14}.

\section{Source sample}\label{sec:srcs}

While several observations of FSRQs with the MAGIC telescopes (\citealt{2008Sci...320.1752M}, \citealt{2021A&A...647A.163M}) resulted in the detection of $\gamma$-ray emission, in this paper, we focus on the observations that have not resulted in a significant detection.

Most of the sources in this study have been observed by MAGIC as a target of opportunity (ToO): OP~313, AO~0235+16, 3C~454.3, TXS~0025+197, B2~22234+28A, B2~0234+28 following alerts of high activity of the sources in other wavelengths by the MWL partners, mainly \textit{Fermi}-LAT. Moreover, 4C+55.17 and TXS 2241+406 were observed within the deep-exposure monitoring program based on their average GeV emission in the preceding years.

\begin{figure}
\includegraphics[scale=0.53]{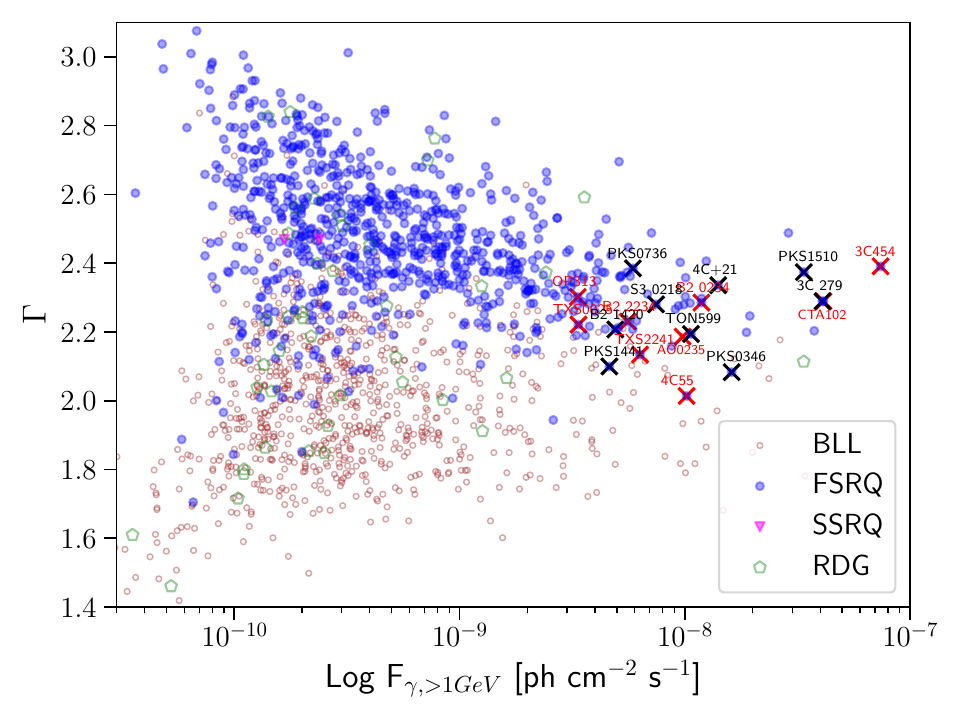}
\caption{Diagram of the gamma-ray flux vs. spectral index for the extragalactic sources with known redshift contained in the 4LAC-DR3 \citep{2022ApJS...263..24A}. FSRQs already detected at VHE gamma-ray and reported in TeVCat \citep{2008ICRC....3.1341W} at VHE gamma-ray are labeled in the plot and reported with black crosses, while the FSRQs investigated in this project are indicated with red crosses.}
\label{fig:plot_lum_index_4lac}
\end{figure}

The sources included in our study are listed in Table~\ref{tab:4fgl_cat} and Table~\ref{tab:magic_obs}. From Fig. \ref{fig:plot_lum_index_4lac}, showing the gamma-ray flux vs. spectral index for the extragalactic sources with known redshift contained in the 4LAC-DR3 \citep{2022ApJS...263..24A}, it is possible to see that the sources selected in this work are among the brightest AGNs in the GeV domain.

Table~\ref{tab:4fgl_cat} contains information about the studied sources based on the information in the 4FGL-DR2 catalog, namely their coordinates, variability index, the integrated flux from the catalog, as well as the values calculated during the data interval when the MAGIC telescopes observed the sources. Table~\ref{tab:magic_obs} provides information about the sources related to the analysis of data from the MAGIC telescopes, such as observation time (exposure), zenith angle range over which the source was observed, date of observations in MJD, the excess signal calculated using the \cite{1983ApJ...272..317L} prescription and the integral ULs.

\begin{table*}
	\centering
	\caption{\textsl{Fermi}-LAT FSRQs properties. Association, 4FGL name and coordinates (in \textsc{J2000} reference frame) are taken from the 4FGL-DR2 catalog, while flux and variability index were obtained by the \textsl{Fermi}-LAT analysis performed in this work. For the selected periods for the flux estimates during MAGIC observation, see Table \ref{tab:magic_obs}. }
	\label{tab:4fgl_cat}
	\begin{tabular}{lcccccccr} 
		\hline
		Association name & 4FGL source name & R.A. & Decl.  & Integral flux (0.1-1000 GeV) & Variability index & Integral flux during MAGIC obs. (0.1-1000 GeV) \\
		& & [$^\circ$] & [$^\circ$] & [$10^{-8}$ cm$^{-2}$ s$^{-1}$] & &  [$10^{-8}$ cm$^{-2}$ s$^{-1}$]\\
		\hline
		TXS 0025+197 & J0028.4+2001 & 7.12   & 20.03  &  1.2  $\pm$ 0.2  & 28    & 63.4 $\pm$ 7.8   \\ 
		B2 0234+28   & J0237.8+2848 & 39.46  & 28.80  &  16.7 $\pm$ 0.4  & 3220  & 80.3 $\pm$ 4.2 &  \\
		AO 0235+16  & J0238.6+1637 & 39.67  & 16.62  &  13.1 $\pm$ 0.5  & 65    & 20.1 $\pm$ 5.1  \\
		4C +55.17    & J0957.6+5523 & 149.42 & 55.38  &  8.5  $\pm$ 0.3  & 33    & 7.3 $\pm$ 1.0   \\
		OP 313       & J1310.5+3221 & 197.65 & 32.35  &  3.7  $\pm$ 0.6  & 170   & 9.0 $\pm$ 5.1 \\
		CTA 102      & J2232.6+1143 & 338.10 & 11.73  &  41.6 $\pm$ 0.6  & 14315  & 1030.0 $\pm$ 20.0 \\
		B2 2234+28A  & J2236.3+2828 & 339.08 & 28.45  &  6.6  $\pm$ 0.3  & 388    & 16.9 $\pm$ 4.8 \\
		TXS 2241+406 & J2244.2+4057 & 341.06 & 40.95  &  6.3  $\pm$ 0.4  & 3930   & 1.9  $\pm$ 1.7 \\
		3C 454.3     & J2253.9+1609 & 343.49 & 16.15  &  215.0 $\pm$ 1.0 & 50905 & 261.0 $\pm$ 8.0\\
		\hline
	\end{tabular}
\end{table*}

\begin{table*}
	\centering
	\caption{Information on data collection by the MAGIC telescopes. For the exact time see Table \ref{tab:magic_fermi_obs}.}
	\label{tab:magic_obs}
	\begin{tabular}{lccc|ccccr} 
		\hline
		Association name & Exposure & Zenith & MJDs - 50000   &  Significance of excess & ULs $E>100$GeV  \\
		& [h] & [$^\circ$] &   & [$\sigma$] & [$10^{-10}$ cm$^{-2}$ s$^{-1}$] & \\
		\hline
		TXS 0025+197 & 5.0  & 9-35 & 8728-8730; 8816-8818  & 0.2 & 3.1\\
		B2 0234+28   & 25.6 & 0-36 & 8379-8481  & 1.6 & 1.3 \\
		AO 0235+16  & 6.1  & 11-26 & 7385-7400 & 0.7 & 1.8\\
		4C +55.17    & 50.0 & 26-42 & 5512-5576, 6671-6777, 6993-7151, 57364  & 1.5 & 1.1\\
		OP 313       & 13.6 & 4-39 & 6774-6811, 8654-8657, 8844-8849  & -0.5 & 2.0  \\
		CTA 102      & 3.2  & 17-42 & 7715-7740; 8105  & 1.7 & 6.0\\
		B2 2234+28A  & 6.7  & 1-47 & 8352, 8639-8677  & 0.5 & 3.1\\
		TXS 2241+406 & 29.5 & 22-35 & 7994, 8665, 8702-8756, 8805-8845 & 0.2 & 3.1 \\
		3C 454.3     & 34.6 & 12-48 & 5505-5509, 6561-6602, 6814-6864, 7257-57260 & 0.6 & 1.7\\
		\hline
	\end{tabular}
\end{table*}

\begin{table*}
	\centering
	\caption{\textsl{Fermi}-LAT FSRQ fit model in the 4FGL catalog and the results of the fit with a logParabola (LP) function (except of 3C 454.3, see text for details) to the \textsl{Fermi}-LAT data simultaneous to the MAGIC observations. The index $b$ in the LAT analysis of 3C454.3 during MAGIC observations has been fixed for fit convergence. }
	\label{tab:fit_fermi}
	\begin{tabular}{lccc|ccccr} 
		\hline
		Association name  & Model &  \multicolumn{2}{c}{\fermi{} fit 4FGL} & \multicolumn{2}{c}{\fermi{} fit during MAGIC obs}  & \\
        & & $\alpha$ & $\beta$ & $\alpha$ & $\beta$ \\  
		\hline
		TXS 0025+197  & LP & 2.092 $\pm$ 0.026 & 0.108 $\pm$ 0.015 & 2.53 $\pm$ 0.24 & 0.41 $\pm$ 0.19\\ 
		B2 0234+28    & LP & 2.27 $\pm$ 0.02 & $0.0898 \pm 0.0091$ & 2.07 $\pm$ 0.06 & 0.10 $\pm$ 0.04\\
		AO 0235+16   & LP & 2.080 $\pm$ 0.018 & 0.0954 $\pm$ 0.0095 & 1.67 $\pm$ 0.17 & 0.21 $\pm$ 0.09  \\
		4C +55.17    &   LP & 1.901 $\pm$ 0.013 &  0.0767 $\pm$ 0.0067 & 1.93 $\pm$ 0.09 & 0.03 $\pm$ 0.04\\
		OP 313       & LP & 2.282 $\pm$ 0.044   &  0.104 $\pm$ 0.0024  & 1.98 $\pm$ 0.23 & 0.26 $\pm$ 0.01 \\
		CTA 102      & LP & 2.261 $\pm$ 0.009   & 0.1007 $\pm$ 0.0060  & 1.95 $\pm$ 0.03 & 0.05 $\pm$ 0.01  \\
		B2 2234+28A  & LP & $2.273 \pm 0.018$   & $0.0898 \pm 0.0091$  & 1.72 $\pm$ 0.23 & 0.06 $\pm$ 0.07 \\
		TXS 2241+406 & LP  & 2.088 $\pm$ 0.025  & $ 0.090 \pm 0.013$   & 2.13 $\pm$ 0.60 & 0.65 $\pm$ 0.56 \\ \hline 
		3C 454.3     & SuperExpPL & $ -\gamma _1 = 2.014 \pm 0.010$  & $b = 0.5183 \pm 0.0066$   & $-\gamma _1 =$-0.69 $\pm$ 0.05 & $b =$ 0.5183\\ 
		\hline
	\end{tabular}
\end{table*}

In agreement with the preferred spectral model reported in the 4FGL-DR2 catalog, most of the sources (except for 3C 454.3) studied in this paper are well described with a log parabola (LP) spectral model $\frac{dN}{dE} \propto E^{-\alpha +\beta \ln(E)} $ in the GeV range. For 3C 454.3, the spectrum is fitted with the power law with a super-exponential cut-off model (PLSuperExpCutOff):  $\frac{dN}{dE} \propto E^{-\gamma _1}  \exp(-(E)^b)$ \footnote{For more information on the spectral model definitions see \url{https://fermi.gsfc.nasa.gov/ssc/data/analysis/scitools/source_models.html}}. The fit parameters are reported in Table~\ref{tab:fit_fermi}.

All sources are classified as FSRQs in the \textit{Fermi}-LAT 10-year Source Catalog (4FGL-DR2), as well as in the latest 4FGL-DR4 (which is based on 14 years of LAT data), except for AO 0235+16 (4FGL J0238.6+1637) which is classified as BL Lac in both data releases (see details below).



These sources were continuously observed by \textit{Fermi}-LAT (see 12 years LCs at Fig.~\ref{fig:plot_LC_Fermi_12year}) and KVA from 2008 to 2020.
We present the integrated flux ranging from 0.1 to 1000 GeV from the \textsl{Fermi}-LAT telescope over 12 years of observations (denoted as \FluxFermiLC) in the fifth column in Table~\ref{tab:4fgl_cat}, and the last column, the integral flux during the MAGIC observations. We compared the different flux states of each source by dividing the \textit{Fermi}-LAT flux estimated during the MAGIC observations by the flux over the whole 12 years (see the following part on the individual source paragraphs).


\subsection{Notes on individual sources}

\begin{figure*}
\includegraphics[scale=0.35]{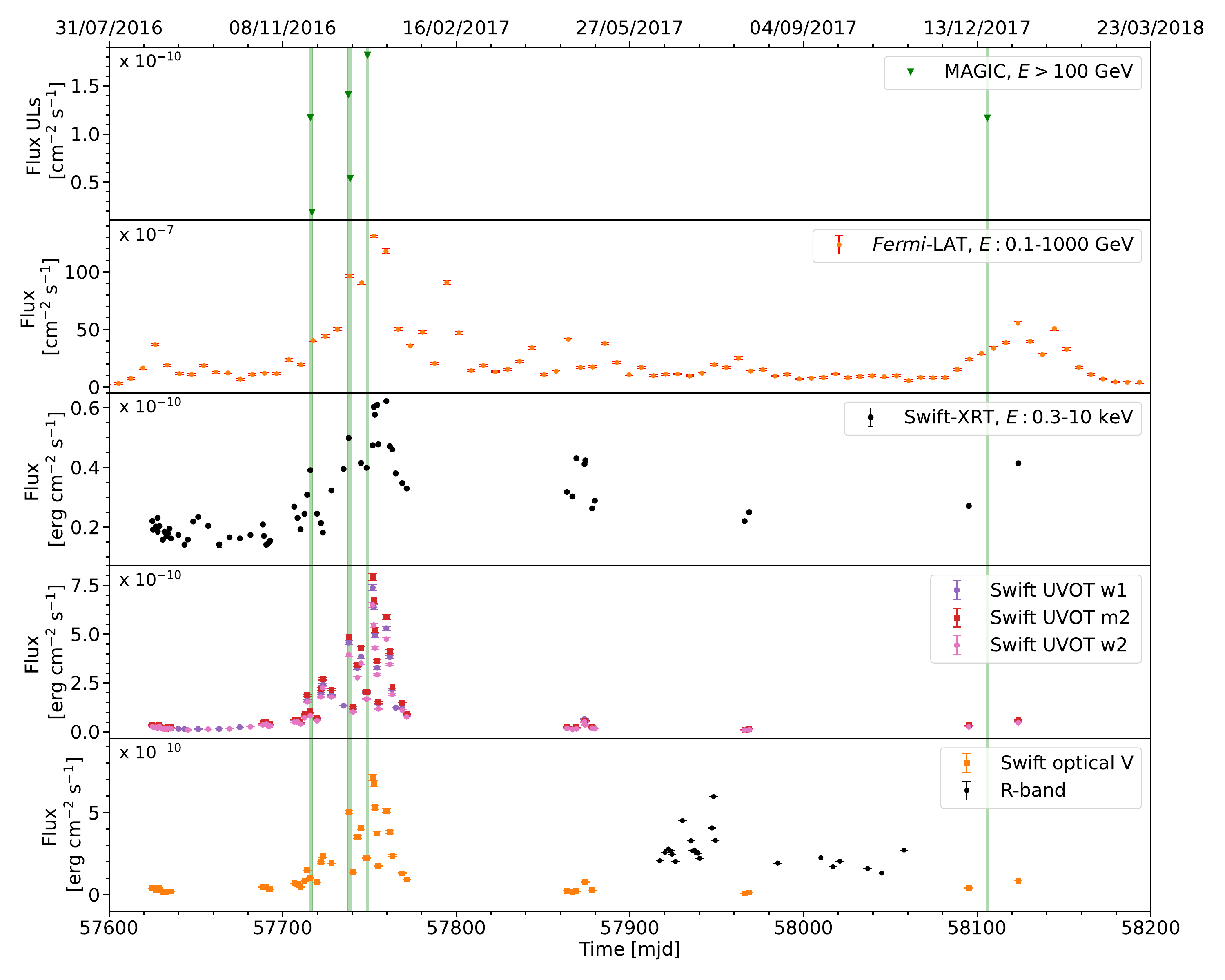}
\caption{Light curve of CTA 102. The green vertical areas indicate the days during which MAGIC observations were carried out. For \textsl{Fermi}-LAT LC, only points that met two criteria: a minimum Test Statistic (TS) value of 9 and a signal-to-noise ratio greater than 2 were selected.}
\label{fig:plot_mwl_cta102}
\end{figure*}

{\bf{CTA 102}} ($z$=1.037, \citealp{1965ApJ...141.1295S}) is one of the most studied FSRQs in the MWL context, but still poorly investigated in the VHE band. High activity in $\gamma$-rays was detected for the first time by the Energetic Gamma Ray Experiment Telescope (EGRET) on board the Compton Gamma-Ray Observatory (\citealp{1993ApJ...414...82N}). CTA~102 is one of the brightest FSRQs observed by \textit{Fermi}-LAT. Strong $\gamma$-ray outbursts have been observed from CTA 102 several times (see \citealp{2019MNRAS.490.5300D}, \citealp{2017Nature...552..374}). From late 2016 to early 2017, CTA~102 exhibited an exceptional outburst that lasted for 4 months, with the fluxes in all bands steadily increasing during the early stage of the high state. As a result, CTA~102 became one of the brightest $\gamma$-ray sources in the sky at that moment \citep{2016ATel.9743....1M, 2016ATel.9756....1C, 2016ATel.9732....1B, 2016ATel.9869....1C}. The MAGIC telescopes followed up CTA 102 during the very high state at the end of 2016 (the flux from \textsl{Fermi}-LAT was 20 -- 30 times stronger than the average \FluxFermiLC ) and also during increasing activity in the HE range at the end of 2017 (the flux from \textsl{Fermi}-LAT was 10 times stronger than \FluxFermiLC ) for a total of $\sim$ 3.5 hours. The CTA 102 optical, UV, and X-ray LCs, demonstrating the source's heightened activity during the period of observation by the MAGIC telescopes, are displayed in Fig.\ref{fig:plot_mwl_cta102}. It also includes the LCs from MAGIC and \textsl{Fermi}-LAT.

{\bf{3C 454.3}} ($z$=0.859, \citealp{2002LEDA.........0P}) is another well studied, highly variable FSRQ. The source was first detected in the GeV range by EGRET \citep{1993ApJ...407L..41H}. 3C 454.3 reached a high flux phase in 2000 and was extremely active in 2005 when it peaked at one of the highest optical brightness recorded from an AGN \citep{2006A&A...453..817V, 2009ApJ...699..817A}. \textit{Fermi}-LAT reported strong and variable $\gamma$-ray emission from this FSRQ in 2008 \citep{2009ApJ...699..817A}. In 2010, during the unusual bright $\gamma$-ray flare, \textit{Fermi}-LAT measured flux at $E >$ 100 MeV  to be (66 $\pm$ 2) $\times$ 10$^{-6}$ photons cm$^{-2}$ s$^{-1}$. This was a factor of three higher than its previous maximum flux recorded in December 2009 \citep{2011ApJ...733L..26A}. At that time, 3C 454.3 was one of the brightest $\gamma$-ray sources in the sky. 
The MAGIC-I telescope observed the source for the first time during the high states of July/August and November/December 2007. The observation was carried out in mono mode. No significant emission was found, and the ULs were derived. 
The obtained data were consistent with the model based on the IC scattering of the ambient photons from the BLR by relativistic electrons, which predicted a sharp cut-off above 20-30 GeV due to the absorption of $\gamma$-rays internally and the reduced effectiveness of the IC emission \citep{2009A&A...498...83A}.
Observations were carried out at different times when the state of the source varied considerably. In November 2010, observations were taken when the source was most active, and the flux was 20 times greater than \FluxFermiLC. In September, October, and November 2013, the source had an average flux at or below \FluxFermiLC. By June and July 2014, the flux had risen to 2 to 4 times higher than \FluxFermiLC. Furthermore, by August 2015, the flux had further increased, roughly 3.5 times greater than \FluxFermiLC. However, MAGIC observations resulted in no significant detection.

These observations, triggered by alerts from multi-wavelength partners such as KVA and \textsl{Fermi}-LAT, emphasizing the time of the MAGIC observations, are depicted in Fig.~\ref{fig:plot_mwl_3C454}. Following data selection, the total effective time of these observations amounted to 32 hours. 

{\bf{OP~313}} ($z=0.997$, \citealp{2010AJ....139.2360S}):  
In 2014, this blazar exhibited an upsurge in its activity in the GeV energy range, which led to its inclusion in the LAT Monitored Sources catalog \citep{2014ATel.6068....1B}. From 2019 onwards, an increase in the source's activity was observed once again, evident in both the optical \citep{2019ATel12898....1B} and the $\gamma$-ray bands \citep{2021ATel14404....1H}. MWL LCs, focusing on the time of the MAGIC observations, are shown in Fig.~\ref{fig:plot_mwl_op313}.
During these periods of high activity, the MAGIC telescopes gathered 12.3 hours of high-quality data. Specifically, in 2014, the flux was documented as 11 times higher than the reference flux, \FluxFermiLC. In 2019, the flux increased to 5 and 10 times that of \FluxFermiLC. Despite these high activity periods, the MAGIC telescopes made no detections at that time. 

In December 2023, LST-1 detected high-energy $\gamma$-ray emissions from OP 313 exhibiting a significant flux level of over 5 $\sigma$, corresponding to 15\% of the Crab Nebula's flux above 100 GeV \citep{2023ATel16381....1C}. MAGIC also observed this  source during that time. However, we do not use that data in this paper. 

{\bf{TXS~0025+197}} ($z=1.552$,  \citealp{2018A&A...613A..51P}) is the FSRQ with the highest redshift among the analyzed sources. 
The \textit{Fermi}-LAT observed an increased $\gamma$-ray flux on 14th August 2019. Preliminary analysis indicates that the source reached a peak daily flux ($E>100$ MeV) of (1.0 $\pm$ 0.2) $\times$ 10$^{-6}$ photons cm$^{-2}$ s$^{-1}$ \citep{2019ATel13032....1B}. MAGIC observed TXS 0025+197 in September and November-December 2019, during an increased activity observed by \textit{Fermi}-LAT in the $\gamma$-ray band (with a flux 50 -- 60 times higher than \FluxFermiLC) and collected 5 hours of good quality data. The \textsl{Fermi}-LAT LC, focusing on the time of the MAGIC observation is shown in Fig~\ref{fig:plot_txs0025}. Unfortunately, simultaneous optical KVA data were not available.

\begin{figure*}
\includegraphics[scale=0.35]{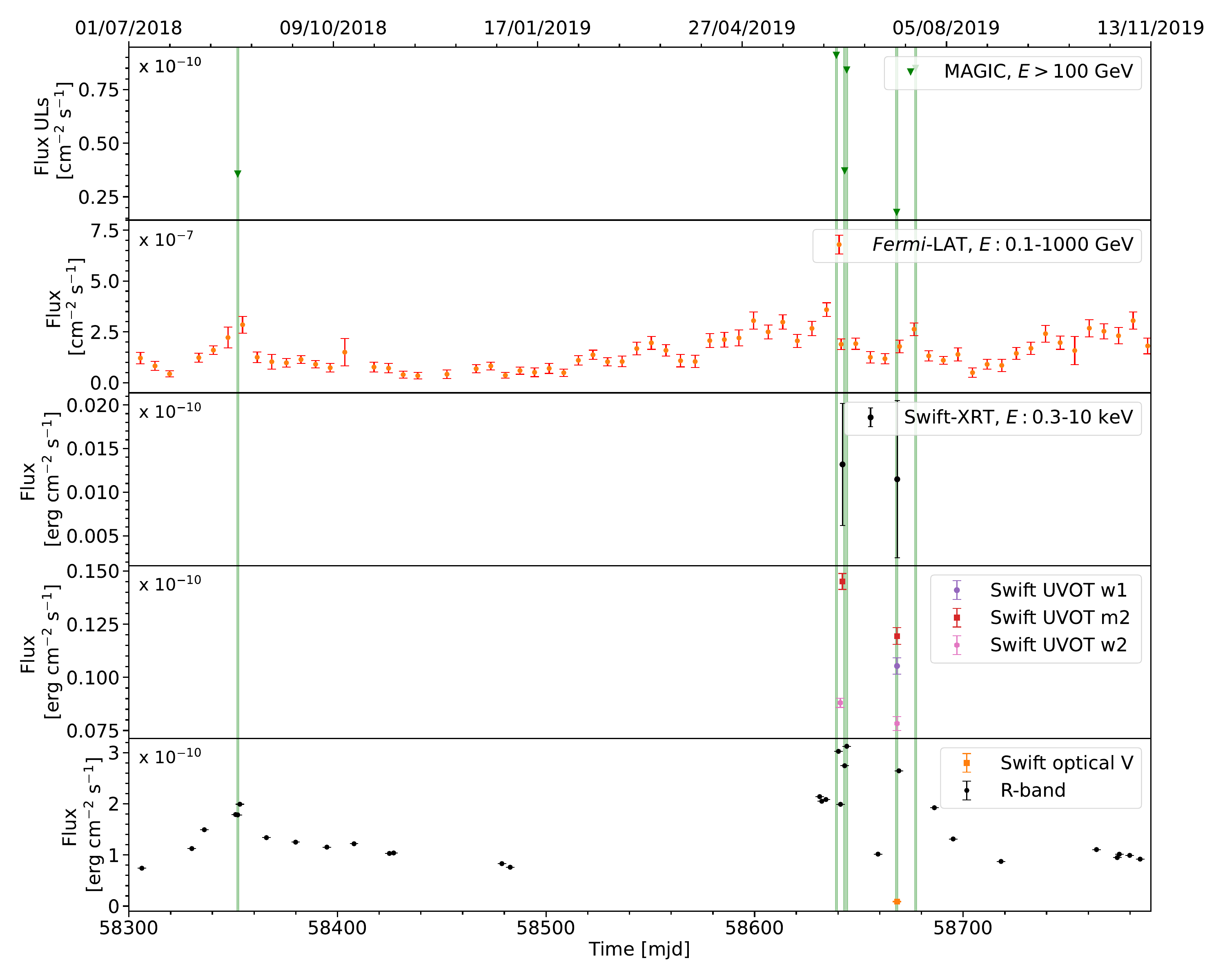}
\caption{Light curve of B2 2234+28A. The green vertical areas indicate the days during which MAGIC observations were carried out. For the \textsl{Fermi}-LAT LC, only points that met two criteria: a minimum Test Statistic (TS) value of 9 and a signal-to-noise ratio greater than 2 are shown.}
\label{fig:plot_mwl_b2_2234}
\end{figure*}

{\bf{B2 2234+28A}} ($z = 0.790$, \citealp{2012ApJ...748...49S}) displayed notable activity that the Guillermo Haro Observatory recorded. In particular, a significant increase in the source's luminosity in the near-infrared (NIR) band was detected. On 26th November 2010, the luminosity of the source in the NIR band increased approximately by a factor of 11 on a daily timescale \citep{2010ATel.3056....1C}. Later in 2016, the same observation revealed a sixfold increase \citep{2016ATel.8572....1C}. MAGIC observed B2 2234+28A during its increased activity in the optical and GeV energy bands observed by KVA and \textit{Fermi}-LAT (flux $\sim$ 1.3 -- 2 times higher than \FluxFermiLC), respectively, in September 2018 and June/July 2019 and collected 6.7 hours of good-quality data. Fig.~\ref{fig:plot_mwl_b2_2234} illustrates the UVOT and XRT light curves, demonstrating that the source was undergoing a phase of amplified activity during the period of MAGIC observations.

{\bf{B2 0234+28}} ($z=1.206$, \citealp{2012ApJ...748...49S}):
In October 2018, the Special Astrophysical Observatory of the Russian Academy of Sciences (SAO RAS) reported a new active phase of the source, which increased its flux in the R band by a factor of 3 magnitudes with respect to its quiet state \citep{2018ATel12111....1V}. The Guillermo Haro Observatory observed a flare in NIR on 5th January 2019. They reported that the source had increased its flux by 50\% \citep{2019ATel12401....1C}. The increase in the flux level happened on a daily timescale. MAGIC observed the source in 2018 and 2019 during its increased activity in the optical band. The source reached the highest flux in October 2018 which was 6--10 times higher than the average flux \FluxFermiLC observed by \textsl{Fermi}-LAT. KVA and \textsl{Fermi}-LAT LCs are shown in Fig.~\ref{fig:plot_mwl_b20234}, covering the time of MAGIC observation. MAGIC followed this source at that time and collected 25.6 h of data.

{\bf{AO 0235+16} } ($z = 0.94$, \citealp{2017AJ....154..114O}): The classification of AO 0235+16 is not certain \citep{2007A&A...464..871R}. It was one of the first objects classified as a BL Lac object \citep{1975ApJ...201..275S} and is still often classified as such. However, it has some characteristics of FSRQs, namely, strong emission lines have been detected in the spectra of A0 0235+16 during faint optical states \citep{1987ApJ...318..577C,1996A&A...314..754N}. The source is also strongly Compton-dominated during the flares, indicating that external seed photons must exist for the Compton scattering \citep{2012ApJ...751..159A}. 
At the end of 2014 and the beginning of 2015, the source showed unusually powerful optical and radio flares \citep{2015ATel.6970....1V, 2015ATel.7004....1S}. AO 0235+16 showed increased activity in the optical band at the end of 2015 and the beginning of 2016, triggering the MAGIC observations (see LC in Fig.~\ref{fig:plot_mwl_ao}). MAGIC collected a total of 6.1\, hours of good-quality data.

{\bf{4C+55.17}} ($z=0.902$, \citealp{2018A&A...613A..51P}) is a bright \textit{Fermi}-LAT FSRQ, which made the source a promising VHE emission candidate, due to high brightness and lack of strong variability (a low variability index is reported in all data releases of the 4FGL catalogs). MAGIC monitored this source in the VHE band (during the low state, flux below average \FluxFermiLC) from November 2010 to January 2011 for 28 hours of good-quality data. No significant VHE $\gamma$-ray signal above 100 GeV was detected. Integral and differential ULs on the $\gamma-$ray flux were derived \citep{2014MNRAS.440..530A}. 
The VERITAS telescope also observed the source for 45 hours between May 2010 and March 2012. These observations also showed no significant VHE $\gamma$-ray signal \citep{2013arXiv1303.1103F}. Between 2008 and 2020, the source state was stable, as can be seen in ~\ref{fig:plot_mwl_4C55}).
In this paper, old and new observations were merged to investigate this source in more detail at the VHE range. To carry out the analysis for this work, data from \cite{2014MNRAS.440..530A} were combined with the MAGIC observations after 2011. After all 80 hours of good-quality data were obtained, including also 50 hours of new data not previously published when the source showed increased activity in the GeV energy range. 

{\bf{TXS 2241+406}} ($z=1.171$, \citealp{2012ApJ...748...49S}): While being a promising candidate to emit VHE gamma rays, it also showed exceptional variability at past times.  For the first time in February 2015, \textit{Fermi}-LAT observed a gamma-ray outburst from this source on a daily timescale  \citep{2015ATel.8319....1O}. During that period, TXS 2241+406 was also monitored with KVA, showing optical variability spanning over 2.5 mag. As can be seen in the MWL LC (Fig.~\ref{fig:plot_txs_2241} and Fig.~\ref{fig:plot_LC_Fermi_12year}), the variability of the source has significantly increased since 2015 compared to the previous six years of \textit{Fermi}-LAT observations, encouraging monitoring with the MAGIC telescopes. In August 2017, MAGIC followed this source for the first time and subsequently conducted a 27-hour observational campaign from July to December 2019. Unfortunately, during this period, the activity of the source was low, and it was either not detected by \textsl{Fermi}-LAT or the flux was below the average \textsl{Fermi}-LAT flux \FluxFermiLC.

\section{Results and discussion}\label{sec:results}
We investigated the MWL behavior of these nine FSRQs contemporaneously with the MAGIC observations. 
We modeled the SED by utilizing data from \textsl{Fermi}-LAT telescope observations while accounting for redshift-dependent absorption by the EBL. Subsequently, we calculated the differential upper limits using data from the MAGIC telescopes, providing insights into the emission properties and the possible additional absorption in the radiation field surrounding the jet, such as the BLR. 

\subsection{Gamma-ray emission}
In Table~\ref{tab:magic_obs}, we report the MAGIC observation results for each of the nine FSRQs. We did not find any statistically significant signal ($>5\sigma$ excess) for any sources in the VHE energy band. The statistically significant excess of $\gamma$-rays was determined using the Li \& Ma formula, as described in \cite{1983ApJ...272..317L}.
We performed simultaneous \textit{Fermi}-LAT SEDs analysis according to the MAGIC observations. We calculated ULs with 95\% C.L. in 5 energy bins in the energy range from 50 GeV to 500 GeV using MAGIC data, with assumed intrinsic spectral index of the $\gamma$-ray photon distribution, $\alpha$=2.2 for all sources. 

The flux for each FSRQ is extrapolated to the VHE range from the \textit{Fermi}-LAT data, considering the absorption of $\gamma$-rays in the EBL as per the \citet{2011MNRAS.410.2556D} model Additionally, more recent EBL models \citep{2017A&A...603A..34F, 2021MNRAS.507.5144S} were also tested showing compatible results. This extrapolation operates under the assumption that there are no breaks in the photon spectra between HE and VHE due to particle distribution cooling. Following this, the extrapolated model is compared to the MAGIC ULs. If the MAGIC upper limits are more constraining than such an extrapolation, it could suggest an absorption-induced cut-off in the VHE range. However, we note that it is possible that different flux states have been combined when the MAGIC and Fermi-LAT data were stacked, which could result in a mismatch in the source spectra between the HE and VHE data. The stacking of the data could, however, average out the HE and VHE gamma-ray data collected during what might have been different flux states.

\begin{figure*}
\includegraphics[scale=0.4]{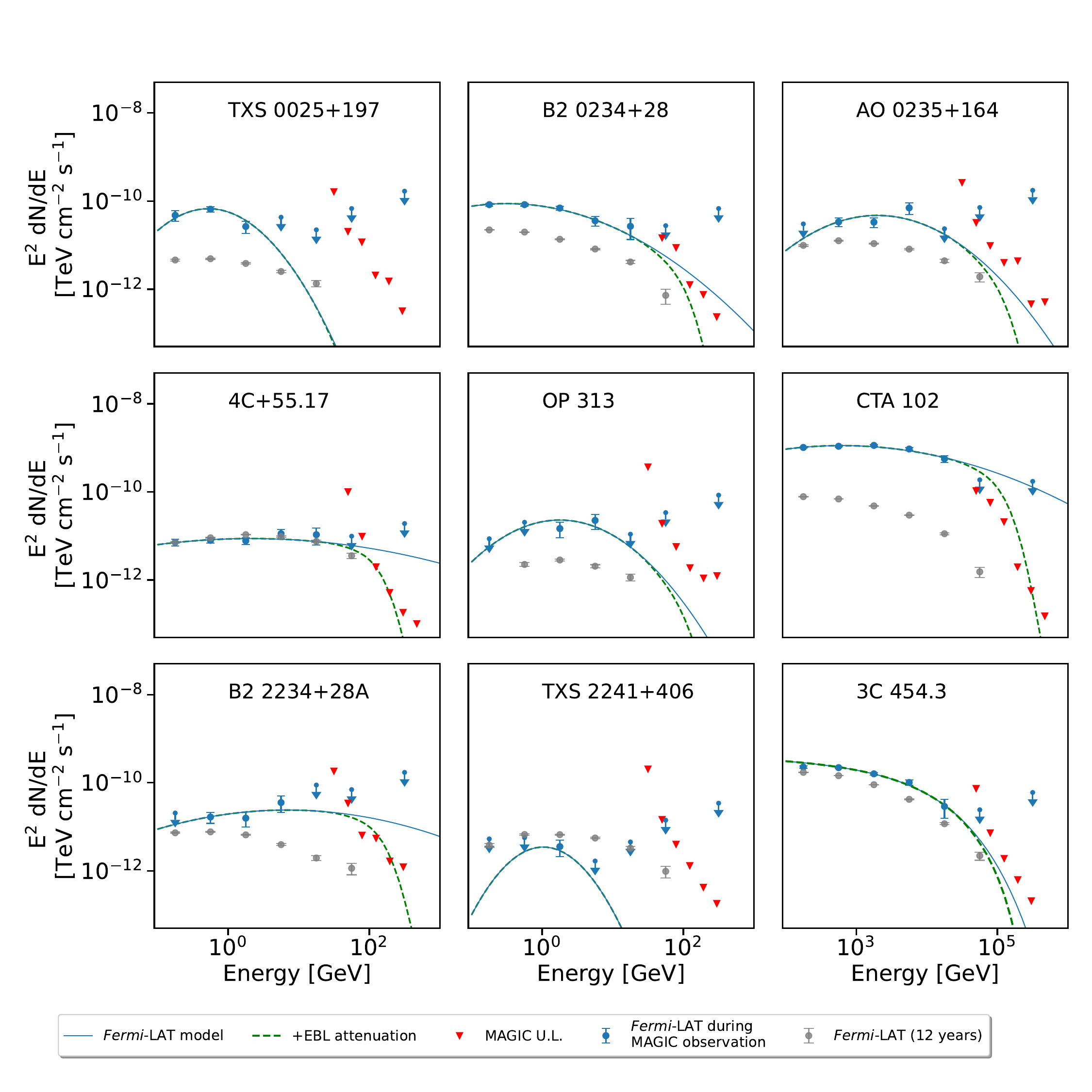}
\caption{SED of all studied sources. The spectra of all studied sources that are simultaneous to MAGIC observations are fitted with log parabolic models (blue solid line) and extrapolated at VHE considering the EBL absorption (green solid line). SED points (gray dots) from data collected by \textit{Fermi}-LAT over a period of 12 years to show the average state of each source are also included in the plots.}
\label{fig:plot_all_sed}
\end{figure*}

The combined results of the \textit{Fermi}-LAT analysis and the ULs from the MAGIC data analysis, along with the HE/VHE SED of all the investigated sources incorporating an EBL attenuation emission model, are presented in Fig.~\ref{fig:plot_all_sed}.

The calculated differential ULs for seven sources are consistent with the LP model extended from \textit{Fermi}-LAT energy attenuated by EBL. Among them, for four sources, the MAGIC ULs lie above the \textit{Fermi}-LAT extrapolated model, and for three of the sources, the ULs are close to the \textit{Fermi}-LAT extrapolated model. 
Regarding the other two sources, B2 2234+28A and CTA~102, the MAGIC ULs around 100 GeV are below the  \textit{Fermi}-LAT EBL extrapolation model.
Therefore, those two objects are plausible candidates for sources in which absorption in BLR could introduce an extra cut-off from absorption in the BLR. We further investigate this possibility in the next section \ref{sec:model} and focus on these two sources. 

\begin{figure*}%
    \centering
    \subfloat{{\includegraphics[scale=0.35]{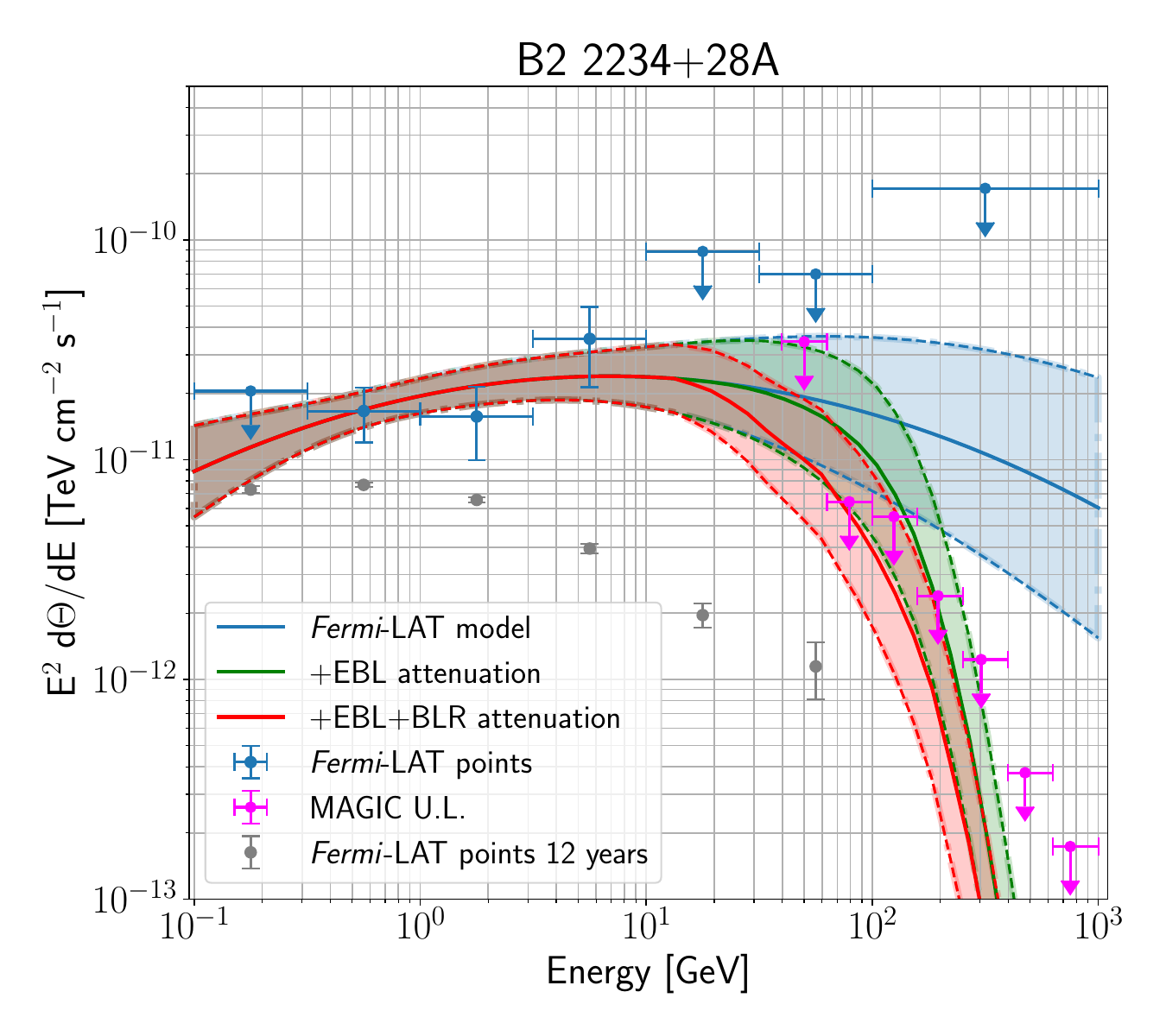}  }}%
    \qquad
    \subfloat{{\includegraphics[scale=0.35]{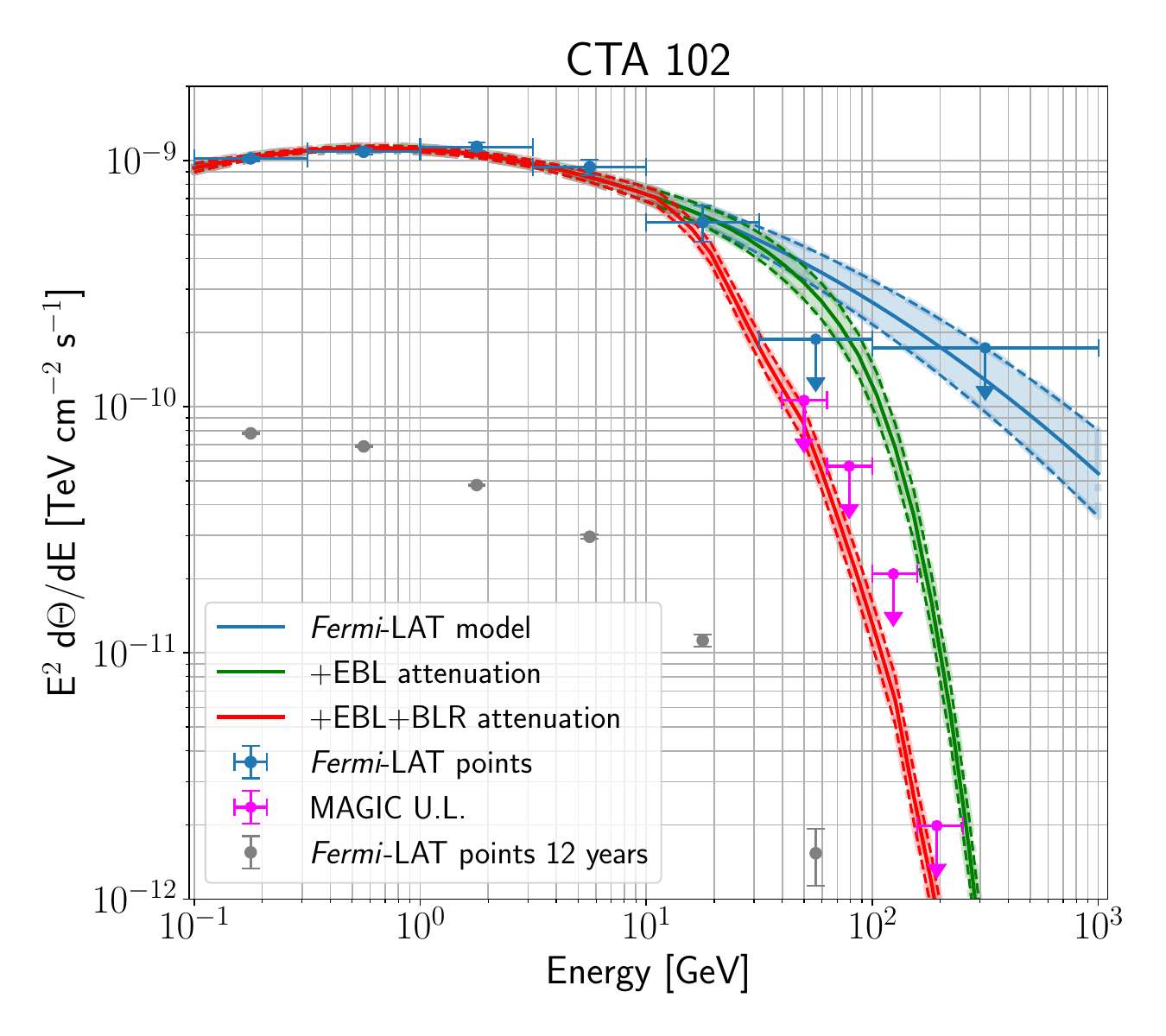} }}%
    \caption{SED of B2 2234+28A (on the left) and CTA 102 (on the right) in the HE and VHE range: derived VHE differential upper limits (95 \% C.L.) on the flux by MAGIC and \textit{Fermi}-LAT spectrum obtained during the MAGIC observation period. A blue solid line depicts a spectral fit with the LP model. The intrinsic \textit{Fermi}-LAT spectrum attenuated with Dominguez's EBL model \citep{2011MNRAS.410.2556D} is shown with a green line. The red line shows the spectrum after considering absorption on multiple BLR lines. The spectra points of B2 2234+28A and CTA 102 obtained in 12 years of \textit{Fermi}-LAT observations are also shown (gray dots). The spectrum is shown taking into account the uncertainty of the parameters obtained when fitting the data (shaded region).}%
    \label{fig:sed_ul_blr}%
\end{figure*}

\subsection{Modelling} \label{sec:model}
As a result of our analyses, two sources, B2 2234+28A and especially CTA 102, showed a hint of cut-off in the HE/VHE SED that cannot be explained only by the EBL absorption. In the case of B2 2234+28A, if we also consider uncertainties with the EBL extrapolation, the EBL absorption might explain the cut-off; in the case of CTA 102, the cut-off is more robust. Subsequently, we investigated the possibility of additional absorption in the BLR for these two sources.

The simple empirical stratified BLR model from \cite{2016ApJ...830...94F} is applied. It relies on the reverberation mapping method of AGNs \citep{2016ASSL..439..249B, 2009ApJ...697..160B}. It assumes that accretion disk radiation is absorbed by the BLR clouds surrounding the emission and is re-emitted as monochromatic lines at an established distance from the emission region. A similar approach was used in a study of 3C 279 \citep{2019A&A...627A.159H}.

Our BLR model comprises 26 concentric infinitesimally thin spherical shells containing gas emitting a range of emission lines, from Ly$\epsilon$ to H$\alpha$. The radius and the luminosity of individual lines of the BLR are required to calculate the $\gamma$-ray absorption and for further broadband SED modeling. 

We estimated the luminosity of the disk, $L_{\text{disk}}$, as well as the luminosity and radius for each of the 26 concentric shells for both sources using the stratified BLR model. We estimated the radius of individual lines in the BLR using the H$\beta$ emission as a reference. All other shell luminosities and radii will be scaled based on the one emitting the H$\beta$ line. To achieve this, we employed the following relation (based on reverberation mapping of AGN objects):
\begin{eqnarray*}
 R(\text{H} \beta) = 10^{16.94 \pm 0.03} \left( \frac{L (5100 \angstrom )}{10^{44} \text{erg s}^{-1} } \right)^{0.533 \pm 0.035} \mathrm{cm}, 
\\ 
\left( \frac{L (5100 \angstrom )}{10^{44} \text{erg s}^{-1} } \right) = \left( \frac{L (\text{H} \beta)} {(1.425 \pm 0.007) \times 10^{42} \text{erg s}^{-1}}  \right)^{0.8826 \pm 0.0039}\\
\end{eqnarray*} as described in \cite{2005ApJ...630..122G} and \cite{2013ApJ...767..149B}.

In our estimation, we used the relative line luminosities $L(\text{MgII})$ from \cite{2021ApJS..253...46P} and converted these values using the broad emission line parameters from \cite{2016ApJ...830...94F}, assuming $L(\text{MgII})/L(\text{H}\beta) = 1.7$.

With the method described above, we derived the following luminosity $L_{H{\beta}}$ and radius $R_{H\beta}$ for line H$\beta$ values for CTA 102: $L_{H\beta}$ = 6.7 $\times
$ 10$^{43}$ erg s$^{-1}$ and $R_{H\beta}$ = 5.13 $\times$ 10$^{17}$ cm, for B2 2234+28A: $L_{H\beta}$ = 1.62 $\times$ 10$^{43}$ erg s$^{-1}$ and $R_{H\beta}$ = 2.67 $\times$ 10$^{17}$ cm. The distances and luminosities of the remaining lines are scaled. To determine the radii and luminosities of other lines, we use the values from Table 5 in \cite{2016ApJ...830...94F}. The absorption in the BLR was calculated over all lines from Table 5 in \cite{2016ApJ...830...94F}.  We calculated the BLR absorption using the \textit{agnpy} modeling package \citep{2022A&A...660A..18N}. Concerning analytical speed and improved numerical accuracy when dealing with the multi-dimensional integration, methods from \textit{cubepy}\footnote{\url{https://github.com/Areustle/cubepy}} were implemented in \textit{agnpy} for those calculations.

The methodology approach to constrain the distance between the black hole and the emission region (a blob) \Rblob and to check its consistency using the broadband model involves two types of models. Firstly, modeling the SED in the HE and VHE ranges (\textit{Fermi}-LAT flux measurements and MAGIC ULs) allows us to estimate \Rblob. This is done using a phenomenological model that constrains the location of the emission region. The absorption in the BLR radiation field is introduced, and the resulting spectrum is compared with the MAGIC ULs on the flux. Secondly, we consider a broadband emission model, which tests the underlying blazar physics and parameters from the phenomenological study. This broadband emission model is used to check the consistency of the previous results obtained using the phenomenological model. The approach we are following involves using the phenomenological model to estimate the distance between the black hole and the emission region, and then using the broadband model to test the underlying physics and parameters to ensure consistency with the previous results.

\subsubsection{Phenomenological model} \label{sub_sec:Pheno_model}
The contemporaneous observations of CTA 102 and B2 2234+28A by \textit{Fermi}-LAT and MAGIC telescopes allowed for a combination of the \textit{Fermi}-LAT spectral fit and the MAGIC ULs to constrain the minimum distance of the emission regions to the black hole, \Rblob. For this purpose, the absorption features caused by pair production of $\gamma$-rays with photons of the BLR in SED are used.  

To constrain the distance of the emission region from the black hole, we used the \textsl{Fermi}-LAT fit model, considering both EBL and BLR absorption. We vary the BLR absorption level by varying the distance of the emission region from the black hole with steps of 0.1 $R_{H\beta}$. By comparing this fit model with EBL and BLR absorption with the measured ULs, we can put constraints on the location of the emission region from the black hole.

Fig.~\ref{fig:sed_ul_blr} shows the $\gamma$-ray SEDs from CTA 102 and B2 2234+28A, 
The two sources for which an additional steepening is caused by internal absorption need to be consistent with the VHE ULs. 

The \textit{absorption} module and the \textit{SphericalShellBLR} geometry from the \textit{agnpy} package were used to construct the phenomenological model. Under the assumption that the steepening/cut-off of the $\gamma$-ray emission in the VHE band is due to the absorption in the BLR, we place a constraint on the maximum distance between the black hole and the emission region \Rblob. For CTA~102 we obtain \Rblob < 1.5 $\times$ $R_{H\beta}$ (where $R_{H\beta}$ = 7.7 $\times 10^{17}$cm) and for B2 2234+28A we got \Rblob < 1.6 $\times$ $R_{H\beta}$ (where $R_{H\beta}$ = 4.3 $\times 10^{17}$cm). It is important to note that the presence of a cut-off in the $\gamma$-ray spectrum at high energies does not always indicate absorption in the BLR. Other factors, such as the cut-off in the emitting particle distribution, can also explain this phenomenon \citep{2018MNRAS.477.4749C}.

\begin{figure}
\includegraphics[scale=0.4]{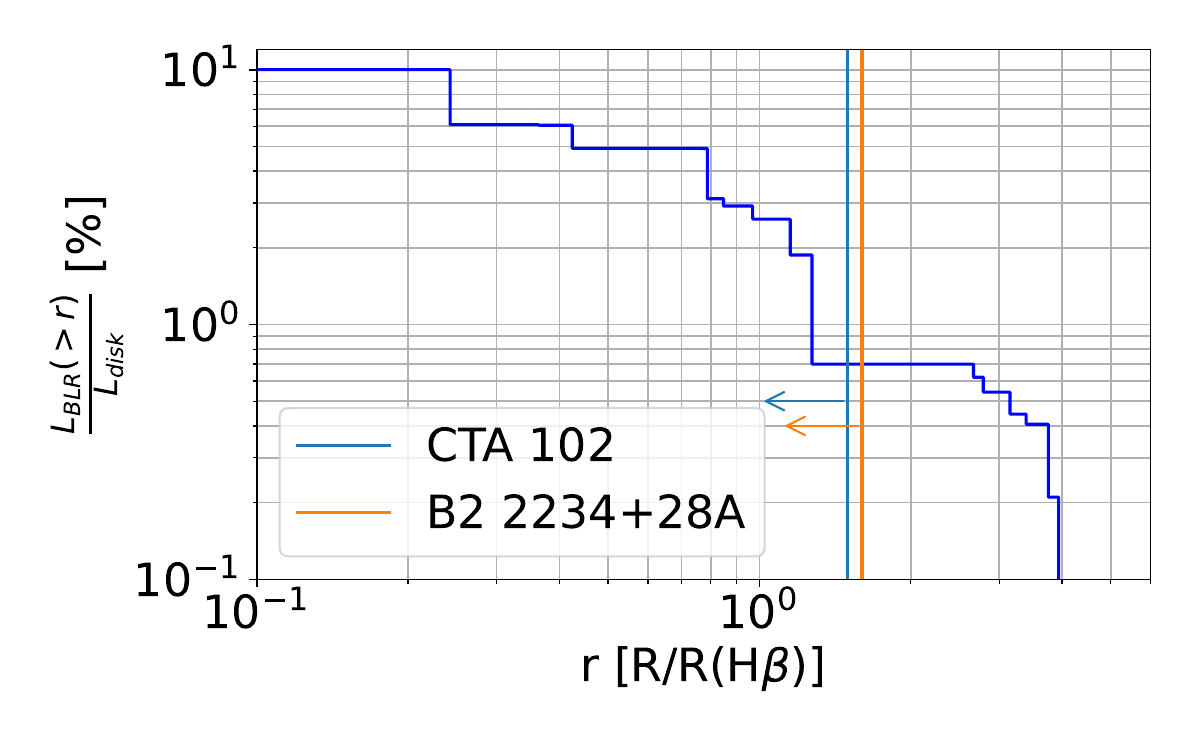}
\caption{
 Fraction (in blue) [\%], $L_{BLR}(>r)/L_{disk}$,  of the disk radiation reprocessed in shells located farther than the distance $r$  from the black hole, according to the used BLR model. 
The distance from the emission region is normalized to the radius of the H$\beta$ line.
Vertical lines show the derived maximum distance from the black hole for the two studied sources to have sufficient BLR absorption to explain the lack of the observed VHE $\gamma$-ray emission. }
\label{fig:plot_disk_radiation}
\end{figure}

The dependency of the integrated disk radiation reprocessed in all the shells located farther than a given distance of the emission region is shown in Fig.~\ref{fig:plot_disk_radiation}. The plot indicates that explaining the non-detection of VHE $\gamma$-ray emission as an effect of the absorption in the BLR requires at least a value as low as \Rblob as the derived value.
The distance is large enough that the emission region is within the outermost part of the BLR. However, even this location provides sufficient absorption to explain the \textit{Fermi}-LAT and MAGIC data, assuming that the emission region is located beyond most of the shells that construct BLR.

\begin{figure*}
    \centering
    \subfloat[\centering B2 2234+28A ]{{\includegraphics[scale=0.7]{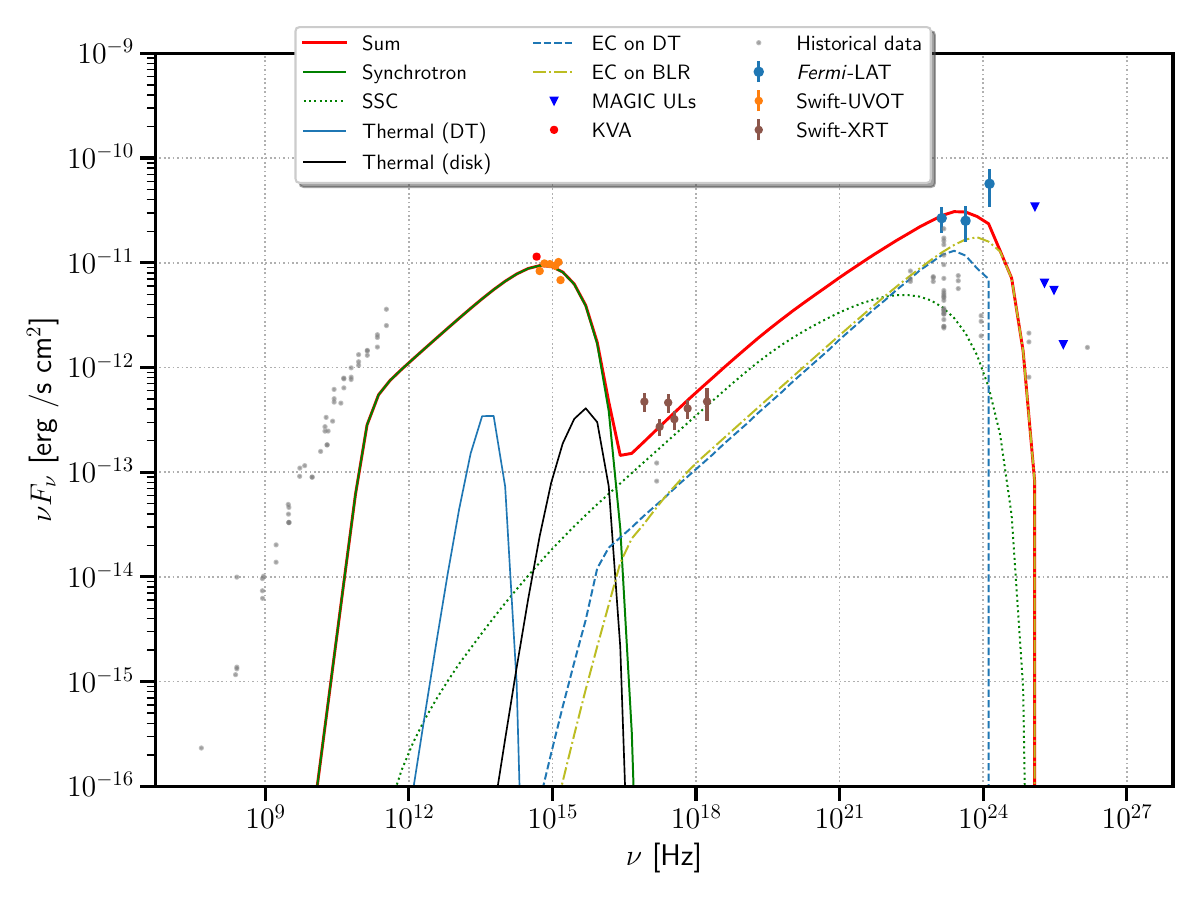}}}%
    \qquad
    \subfloat[\centering CTA 102 ]{{\includegraphics[scale=0.7]{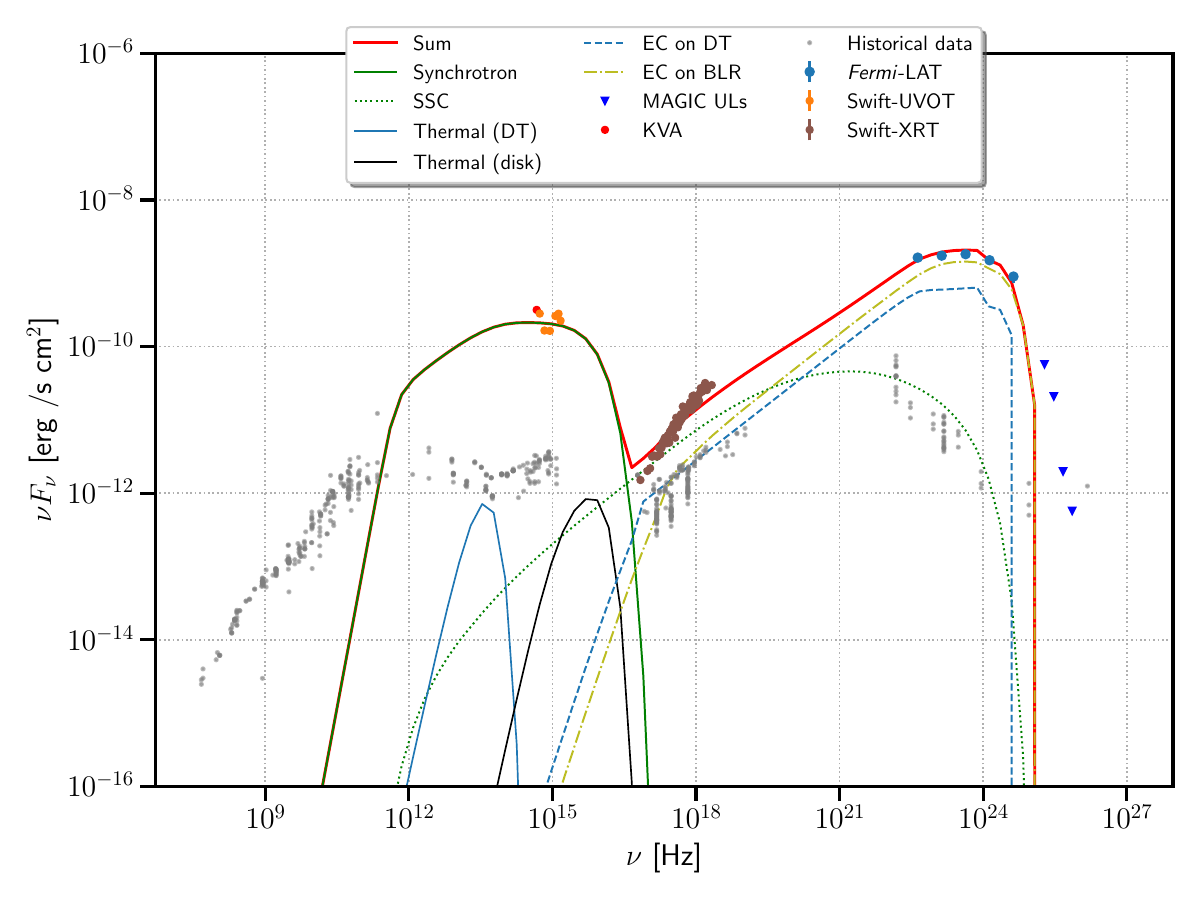} }}%
    \caption{Broadband modeling with agnpy of B2 2234+28A (top panel) and CTA 102 (bottom panel). The solid red line shows the overall emission modeled. The low-energy peak is dominated by synchrotron radiation (green solid line), and the high-energy peak is dominated by the emission produced in the external Compton mechanism using the seed photons from infrared dusty torus (blue dashed line) and broad line region produced in 26 shells (yellow dash-dot line). Gray points -- archival data from the Space Science Data Center - ASI. }%
    \label{fig:sed_modeling}%
\end{figure*}

\subsubsection{Broad-band modeling}
In the previous section, we determined the distance between the emission region and the black hole, denoted as \Rblob, in a phenomenological way based on observations made by the \textsl{Fermi}-LAT and MAGIC telescopes. We now compare these obtained values with a physical model that describes the broadband emission of the jet. 

One-zone SSC models, along with EC models, are typically utilized to model the $\gamma$-ray emission from FSRQs. In the context of FSRQs, it is generally assumed that the central engine is surrounded by clouds re-scattering emission from the accretion disk. Additionally, a disk torus (DT) surrounds the BLR. The radiation field of the accretion disk, reprocessed in the BLR and DT, provides additional targets for the IC process and $\gamma$-ray pair production absorption. However, such a simplified, homogeneous, one-zone leptonic model may not fully represent the complex conditions within the jet. Moreover, detecting VHE $\gamma$-rays from a few FSRQs complicates this simple model, suggesting that the emission region cannot be deeply located within the BLR since the VHE $\gamma$-rays would be absorbed. Consequently, the emission regions are expected to be located comparatively farther from the black hole \citep{2018heas.confE..33Z}. 

Given that our source does not exhibit significant VHE photon emission, the most straightforward target for the EC model would be the BLR radiation field, which strongly absorbs VHE $\gamma$-rays in the sub-TeV range. We assembled data from KVA, \textit{Swift}-UVOT, \textit{Swift}-XRT, \textit{Fermi}-LAT and ULs from the MAGIC telescopes to construct broadband SEDs of B2 2234+28A and CTA 102. 
Both sources are modeled in a framework of a simple one-zone leptonic model with \textit{agnpy}, in which a spherical emission region with a radius $R_{blob}$ is isotropically and homogeneously filled with a magnetic field $B$ and electrons. The electron spectrum spans from $\gamma_{min}$ to $\gamma_{max}$ with indices $p_1$ and $p_2$ below and above the break at $\gamma_b$ and is described by a broken power-law electron energy distribution with spectral normalization $k_e$, differential number density at $\gamma_b$. The single emission region moves along the jet with the bulk Lorentz factor $\Gamma$ at an angle $\theta$ to the line of sight, causing the corresponding Doppler factor $\delta_D$. In our model, we assume that the emission region is situated at a distance that explains the steepening of emission due to the absorption as a result of the phenomenological study (see subsection \ref{sub_sec:Pheno_model}).

\begin{table}
	\centering
	\caption{Parameters used for modeling sources with agnpy, utilizing an emission region represented as a blob, DT, and BLR. Parameters such as $K_e$, $p1$, $\gamma_b$, $\gamma_{max}$, $B$ were derived during the fitting process. $R_{blob}$ was estimated using the phenomenological model. The remaining parameters were fixed according to the information gathered from the literature.}
	\label{tab:fit_parameters_table}
	\begin{tabular}{lcccccc} 
		\hline
            parameter & CTA 102 & B2 2234+28A \\
		\hline
	   Mass [10$^{8}$ $M_\odot\ $] & 12.30 & 2.75 \\	
          \Rblob < [$R_{H\beta}$] & 1.5  & 1.6 \\ \hline 
           & emission regions &  \\ \hline
          $p1$              & 1.97 $\pm$ 0.02 & 2.18 $\pm$ 0.08 \\
          $p2$              & 2.97 & 3.18 \\
          $\delta_D$        &  40  & 13\\
          $R_{blob}$ [10$^{15}$ cm] & 28.5  & 20.8 \\ 
          $K_e $ [10$^{-5}$ cm$^{-3}$]& 546 $\pm$ 24    & 4.47 $\pm$ 0.05 \\ 
          $\Gamma$           & 20.5     & 7\\
          $\gamma_b$         &  850 $\pm$ 2  & 5929 $\pm$ 271 \\
          $\gamma_{min}$     &   1   & 1 \\
          $\gamma_{max}$     & 8616 $\pm$ 19  & 16 292 $\pm$ 4021 \\ 
          $ B $ [Gauss]      & 1.12 $\pm$ 0.04 & 0.52 $\pm$ 0.03 \\ \hline 
           & Dusty Torus &  \\ \hline
          $\xi_{dt}$ & 0.6 & 0.6 \\
          $T_{dt}$ [K] & 1000 & 1000 \\
          $R_{dt}$ 10$^{17}$ [cm] & 159 & 78 \\ \hline
          & Broad-Line Region & \\ \hline
          $R_{H\beta}$ [10$^{17}$ cm]     &   5.13      &   2.67  \\
          $L_{H\beta}$ [10$^{43}$ erg /s]    & 6.7  & 1.62 \\
          $\xi $ & 0.1 & 0.1 \\ \hline
          $U_e$ [erg /cm$^3$]   & 0.022   & 0.027  \\ 
          $U_B$  [erg /cm$^3$]  & 0.047   & 0.011\\
 	\end{tabular}
\end{table}

In our model, the radiation field consists of multiple lines from the corresponding BLR shells and also the thermal IR radiation from the DT.  
The DT is assumed to be an infinitesimally thin ring with temperature $T_{dt} = 10^3$ K \cite{2016ApJ...830...94F}, and the radius of the ring representing the torus $R_{dt}$ is estimated from equation 96 from \cite{2016ApJ...830...94F}. The model also includes a synchrotron (taking into account the synchrotron self-absorption) and an SSC process, which are responsible for the emission in the radio to X-ray energy range.   

Our procedure ensures that the DT emission does not surpasses the synchrotron radiation. While the contribution of DT thermal emission to the SED of the MWL is typically considered for FSRQs, it is often overshadowed by the dominant synchrotron emission. We use the single-temperature black-body radiation computed with \textit{agnpy} to assess this. It allows us to determine if the DT emission is significantly lower in magnitude compared to the synchrotron emission.


Our model assumes that the emission region is outside most of the shells in the BLR, as shown in Fig. \ref{fig:plot_disk_radiation}. The radius of the emission region is established according to the formula $R_{blob} = R\_{blob}\_{BH} /\ \Gamma $, i.e., assuming a conical jet with a half-opening angle of $\sim 1/ \Gamma$.

For CTA 102, we selected a Doppler factor $\delta$ of 40, based on VLBI studies, which established it at 34$\pm$4 to explain HE emissions \citep{2019A&A...622A.158C}. For B2 2234+28A, we used $\delta$ = 13 as a weighted average Doppler factor value for the FSRQ from \textsl{VLBA-BU-
BLAZAR} study \citep{2017ApJ...846...98J}.

The model parameters ($K_e$, $p_1$, $\gamma_b$, $\gamma_{max}$, $B$) were estimated by fitting the SED with the open source package \textit{gammapy} \citep{2017ICRC...35..766D} using a \textsl{Synchtron}, \textsl{SynchrotronSelfComtpon}, and \textsl{ExternalCompton} modules from \textit{agnpy}. We assume a classical cooling break setting $p_2$=$p_1$+1. The fitting procedure was performed, taking into account a simplified systematic error on the flux points. 
We use a conservative estimation of the systematic errors, i.e., 10\% for the HE and X-ray instruments and 5\% for the optical telescopes. The result of the fitting is shown in Fig~\ref{fig:sed_modeling}, while the parameters are given in Table~\ref{tab:fit_parameters_table}. The absorption processes in both EBL and BLRs affect the modeling and interpretation of the data. The modeling results are corrected by considering the absorption on those two radiation fields (red solid line).
 We had a constraint that distance between the emission region and the black hole had to be $< 1.5 \times R_{H \beta}$ for CTA~102, and $< 1.6 \times R_{H \beta}$ for B2 2234+28A from the previous section. Broadband modeling with the leptonic model performed in this section indicates that we may be able to reconstruct the observations with constraints from the phenomenological model.

\section{Summary and Conclusions}\label{sec:summary}
This paper presents a catalog of upper limits for nine FSRQs observed by the MAGIC telescopes over the last 10+ years, resulting in a total observation time of 174 hours for all the sources. All investigated sources are at large redshifts (z $\sim$ 1). 

We compared the limits on the VHE $\gamma$-ray emission of these sources derived with the MAGIC telescopes with the extrapolation of the contemporaneous GeV emission seen by \textsl{Fermi}-LAT, taking into account the absorption by the EBL. For four out of nine investigated sources (namely TXS 0025+197, AO 0235+164, OP 313, and TXS 2241+406), the MAGIC telescopes ULs lie above the emission predicted by the model, which was constructed from the extrapolation of the \textit{Fermi}-LAT observations. For the other three sources (4C +55.17, 3C 454.3, and B2 0234+28), the spectra, after accounting for absorption in the EBL, are close to the ULs obtained by MAGIC (see Fig.~\ref{fig:plot_all_sed}). The fact that for these seven sources, the MAGIC ULs lie very close to or above the \textit{Fermi}-LAT extrapolated model does not allow us to set any additional constraints on the absorption by the BLR. Their large redshift distances could explain the lack of detection in the VHE range for these sources. 
It may also be due to the fact that there is a certain delay between the emission enhancement triggering the ToO and the time when pointing instruments, such as the MAGIC telescopes, start their observations, which can also be additionally limited by atmospheric conditions or moonlight. 

Lastly, for two sources, B2 2234+28A and CTA 102, we obtained with the MAGIC telescopes ULs on the flux below the emission predicted by the \textsl{Fermi}-LAT extrapolation model, which could suggest the presence of an additional absorption from the BLR. As shown in Fig.~\ref{fig:plot_disk_radiation}, the required absorption in BLR to explain the constraints derived by the MAGIC telescopes for both sources requires a weak absorption of only ~1\% of the disk luminosity corresponding to the distance between the emission region and the black hole in both cases at the edge of the BLR, namely $\lesssim 1.6 \times R_{H\beta}$. This agrees with findings for another FSRQ object in \citet{2021ApJ...917...32W}, which suggest that the $\gamma$-ray emission located in 3C 279 most likely originates from the edge of the BLR. 

We investigated two approaches: a phenomenological description of the $\gamma$-ray band spectral shape and a fitting of a broadband radiative model. The first was limited to the HE and VHE $\gamma-$ray range and was used to derive a constraint on the distance between the emission region and the black hole.
The second approach instead considers the emission over the whole spectrum, where the constraint from the phenomenological approach was tested in a leptonic emission model. Based on the SED shown in Fig.~\ref{fig:sed_modeling}, it can be said that the major contribution to the HE emissions from the two sources, B2 2234+28A and CTA 102, studied in this work is the combination of the EC processes on the DT and the BLRs. The data fitting process yielded the values of parameters that describe the broad-band emission model.
For CTA102 and B2 2234+28A, the energy density of the magnetic field ($U_B$) and the total energy density of electrons ($U_e$) appear similar.

It is important to note that the conclusion drawn is highly dependent on the model used. Based on our phenomenological model, our analysis of the observations made on CTA 102 and B2 2234+28A in the HE and VHE ranges leads us to conclude that the absorption in the BLR is weak. Furthermore, the model's best-fit solution, which considers the full broadband spectrum, is consistent with the ULs set by MAGIC. This consistency can be attributed to the limited energy range of the electrons. The broadband modeling, when considering the assumed distance of the emission region, agrees with the observations from the MAGIC telescopes. Therefore, we can infer that the observed steepening at VHE energies is primarily due to the characteristics of the particle distribution, such as its maximum energy or distribution slope, rather than being significantly influenced by absorption effects.
These observations are consistent with the study performed with the \textsl{Fermi}-LAT telescope, which found no evidence for the expected BLR absorption \citep{2018MNRAS.477.4749C}. It is crucial to note that location constraints based on a naive extrapolation of the \textit{Fermi}-LAT spectrum may not be robust. This is because we cannot assume the intrinsic spectrum to behave straightforwardly, which should be considered in future studies, such as those with Cherenkov Telescope Array (CTA, \citealp{2019ICRC...36..741M}) data. The lack of evidence for strong absorption of the VHE $\gamma$-ray radiation in FSRQs is promising for future observations with the present and next generation of IACTs.

\section*{Acknowledgements}
P. Gliwny: project leadership, MAGIC data analysis, theoretical interpretation, paper drafting and editing; H. A. Mondal: MAGIC analysis cross-check, paper drafting and editing; G. Principe: \textit{Fermi}-LAT analysis, paper drafting and editing. Elina Lindfors: KVA analysis and paper editing; F. Longo: supervising the \textit{Fermi}-LAT analysis; P. Majumdar: supervising the MAGIC analysis cross-check; J. Sitarek: paper drafting and theoretical interpretation; N. Zywucka: Swift analysis.

The rest of the authors have contributed in one or several of the following ways: design, construction, maintenance and operation of the instrument(s) used to acquire the data; preparation and/or evaluation of the observation proposals; data acquisition, processing, calibration and/or reduction; production of analysis tools and/or related Monte-Carlo simulations; discussion and approval of the contents of the draft.

GP acknowledges support by ICSC – Centro Nazionale di Ricerca in High Performance Computing, Big Data and Quantum Computing, funded by European Union – NextGenerationEU.

The \textit{Fermi} LAT Collaboration acknowledges generous ongoing support from a number of agencies and institutes that have supported both the development and the operation of the LAT as well as scientific data analysis. These include the National Aeronautics and Space Administration and the Department of Energy in the United States, the Commissariat à l'Energie Atomique and the Centre National de la Recherche Scientifique / Institut National de Physique Nucléaire et de Physique des Particules in France, the Agenzia Spaziale Italiana and the Istituto Nazionale di Fisica Nucleare in Italy, the Ministry of Education, Culture, Sports, Science and Technology (MEXT), High Energy Accelerator Research Organization (KEK) and Japan Aerospace Exploration Agency (JAXA) in Japan, and the K. A. Wallenberg Foundation, the Swedish Research Council and the Swedish National Space Board in Sweden. 
Additional support for science analysis during the operations phase is gratefully acknowledged from the Istituto Nazionale di Astrofisica in Italy and the Centre National d'Etudes Spatiales in France. This work is performed in part under DOE Contract DE-AC02-76SF00515.

We would like to thank the Instituto de Astrof\'{\i}sica de Canarias for the excellent working conditions at the Observatorio del Roque de los Muchachos in La Palma. The financial support of the German BMBF, MPG and HGF; the Italian INFN and INAF; the Swiss National Fund SNF; the grants PID2019-104114RB-C31, PID2019-104114RB-C32, PID2019-104114RB-C33, PID2019-105510GB-C31, PID2019-107847RB-C41, PID2019-107847RB-C42, PID2019-107847RB-C44, PID2019-107988GB-C22, PID2022-136828NB-C41, PID2022-137810NB-C22, PID2022-138172NB-C41, PID2022-138172NB-C42, PID2022-138172NB-C43, PID2022-139117NB-C41, PID2022-139117NB-C42, PID2022-139117NB-C43, PID2022-139117NB-C44 funded by the Spanish MCIN/AEI/ 10.13039/501100011033 and “ERDF A way of making Europe”; the Indian Department of Atomic Energy; the Japanese ICRR, the University of Tokyo, JSPS, and MEXT; the Bulgarian Ministry of Education and Science, National RI Roadmap Project DO1-400/18.12.2020 and the Academy of Finland grant nr. 320045 is gratefully acknowledged. This work was also been supported by Centros de Excelencia ``Severo Ochoa'' y Unidades ``Mar\'{\i}a de Maeztu'' program of the Spanish MCIN/AEI/ 10.13039/501100011033 (CEX2019-000920-S, CEX2019-000918-M, CEX2021-001131-S) and by the CERCA institution and grants 2021SGR00426 and 2021SGR00773 of the Generalitat de Catalunya; by the Croatian Science Foundation (HrZZ) Project IP-2022-10-4595 and the University of Rijeka Project uniri-prirod-18-48; by the Deutsche Forschungsgemeinschaft (SFB1491) and by the Lamarr-Institute for Machine Learning and Artificial Intelligence; by the Polish Ministry Of Education and Science grant No. 2021/WK/08; and by the Brazilian MCTIC, CNPq and FAPERJ.





\bibliographystyle{mnras}
\bibliography{example} 



\newpage
\appendix
\section*{Affiliations}
$^{1}$ {Japanese MAGIC Group: Institute for Cosmic Ray Research (ICRR), The University of Tokyo, Kashiwa, 277-8582 Chiba, Japan} \\
$^{2}$ {ETH Z\"urich, CH-8093 Z\"urich, Switzerland} \\
$^{3}$ {Universit\`a di Siena and INFN Pisa, I-53100 Siena, Italy} \\
$^{4}$ {Institut de F\'isica d'Altes Energies (IFAE), The Barcelona Institute of Science and Technology (BIST), E-08193 Bellaterra (Barcelona), Spain} \\
$^{5}$ {Universitat de Barcelona, ICCUB, IEEC-UB, E-08028 Barcelona, Spain} \\
$^{6}$ {Instituto de Astrof\'isica de Andaluc\'ia-CSIC, Glorieta de la Astronom\'ia s/n, 18008, Granada, Spain} \\
$^{7}$ {National Institute for Astrophysics (INAF), I-00136 Rome, Italy} \\
$^{8}$ {Universit\`a di Udine and INFN Trieste, I-33100 Udine, Italy} \\
$^{9}$ {Max-Planck-Institut f\"ur Physik, D-85748 Garching, Germany} \\
$^{10}$ {Universit\`a di Padova and INFN, I-35131 Padova, Italy} \\
$^{11}$ {Croatian MAGIC Group: University of Zagreb, Faculty of Electrical Engineering and Computing (FER), 10000 Zagreb, Croatia} \\
$^{12}$ {IPARCOS Institute and EMFTEL Department, Universidad Complutense de Madrid, E-28040 Madrid, Spain} \\
$^{13}$ {Centro Brasileiro de Pesquisas F\'isicas (CBPF), 22290-180 URCA, Rio de Janeiro (RJ), Brazil} \\
$^{14}$ {Instituto de Astrof\'isica de Canarias and Dpto. de  Astrof\'isica, Universidad de La Laguna, E-38200, La Laguna, Tenerife, Spain} \\
$^{15}$ {University of Lodz, Faculty of Physics and Applied Informatics, Department of Astrophysics, 90-236 Lodz, Poland} \\
$^{16}$ {Centro de Investigaciones Energ\'eticas, Medioambientales y Tecnol\'ogicas, E-28040 Madrid, Spain} \\
$^{17}$ {Departament de F\'isica, and CERES-IEEC, Universitat Aut\`onoma de Barcelona, E-08193 Bellaterra, Spain} \\
$^{18}$ {Universit\`a di Pisa and INFN Pisa, I-56126 Pisa, Italy} \\
$^{19}$ {INFN MAGIC Group: INFN Sezione di Bari and Dipartimento Interateneo di Fisica dell'Universit\`a e del Politecnico di Bari, I-70125 Bari, Italy} \\
$^{20}$ {Department for Physics and Technology, University of Bergen, Norway} \\
$^{21}$ {INFN MAGIC Group: INFN Sezione di Torino and Universit\`a degli Studi di Torino, I-10125 Torino, Italy} \\
$^{23}$ {Universit\"at W\"urzburg, D-97074 W\"urzburg, Germany} \\
$^{24}$ {Technische Universit\"at Dortmund, D-44221 Dortmund, Germany} \\
$^{25}$ {Japanese MAGIC Group: Physics Program, Graduate School of Advanced Science and Engineering, Hiroshima University, 739-8526 Hiroshima, Japan} \\
$^{26}$ {Deutsches Elektronen-Synchrotron (DESY), D-15738 Zeuthen, Germany} \\
$^{27}$ {Armenian MAGIC Group: ICRANet-Armenia, 0019 Yerevan, Armenia} \\
$^{28}$ {Croatian MAGIC Group: Josip Juraj Strossmayer University of Osijek, Department of Physics, 31000 Osijek, Croatia} \\
$^{29}$ {Finnish MAGIC Group: Finnish Centre for Astronomy with ESO, Department of Physics and Astronomy, University of Turku, FI-20014 Turku, Finland} \\
$^{30}$ {Japanese MAGIC Group: Department of Physics, Tokai University, Hiratsuka, 259-1292 Kanagawa, Japan} \\
$^{31}$ {University of Geneva, Chemin d'Ecogia 16, CH-1290 Versoix, Switzerland} \\
$^{32}$ {Saha Institute of Nuclear Physics, A CI of Homi Bhabha National Institute, Kolkata 700064, West Bengal, India} \\
$^{33}$ {Inst. for Nucl. Research and Nucl. Energy, Bulgarian Academy of Sciences, BG-1784 Sofia, Bulgaria} \\
$^{34}$ {Croatian MAGIC Group: University of Rijeka, Faculty of Physics, 51000 Rijeka, Croatia} \\
$^{35}$ {Japanese MAGIC Group: Department of Physics, Yamagata University, Yamagata 990-8560, Japan} \\
$^{36}$ {Finnish MAGIC Group: Space Physics and Astronomy Research Unit, University of Oulu, FI-90014 Oulu, Finland} \\
$^{37}$ {Japanese MAGIC Group: Chiba University, ICEHAP, 263-8522 Chiba, Japan} \\
$^{38}$ {Japanese MAGIC Group: Institute for Space-Earth Environmental Research and Kobayashi-Maskawa Institute for the Origin of Particles and the Universe, Nagoya University, 464-6801 Nagoya, Japan} \\
$^{39}$ {Japanese MAGIC Group: Department of Physics, Kyoto University, 606-8502 Kyoto, Japan} \\
$^{40}$ {INFN MAGIC Group: INFN Roma Tor Vergata, I-00133 Roma, Italy} \\
$^{41}$ {Japanese MAGIC Group: Department of Physics, Konan University, Kobe, Hyogo 658-8501, Japan} \\
$^{42}$ {also at International Center for Relativistic Astrophysics (ICRA), Rome, Italy} \\
$^{43}$ {also at Port d'Informaci\'o Cient\'ifica (PIC), E-08193 Bellaterra (Barcelona), Spain} \\
$^{44}$ {also at Dipartimento di Fisica, Universit\`a di Trieste, I-34127 Trieste, Italy} \\
$^{45}$ {Max-Planck-Institut f\"ur Physik, D-85748 Garching, Germany} \\
$^{46}$ {also at INAF Padova} \\
$^{47}$ {Japanese MAGIC Group: Institute for Cosmic Ray Research (ICRR), The University of Tokyo, Kashiwa, 277-8582 Chiba, Japan} 
$^{48}$ {Space Science Data Center - Agenzia Spaziale Italiana, Via del Politecnico, snc, I-00133, Roma, Italy}
$^{49}$ {INAF - Istituto di Radioastronomia, Via Gobetti 101, 40129 Bologna, Italy}
\section{Light curves of the remaining sources}
In this part of the paper, we present plots that are not included in the main text of the paper.

\subsection{Light curves of all sources}

\begin{figure*}
\includegraphics[scale=0.5]{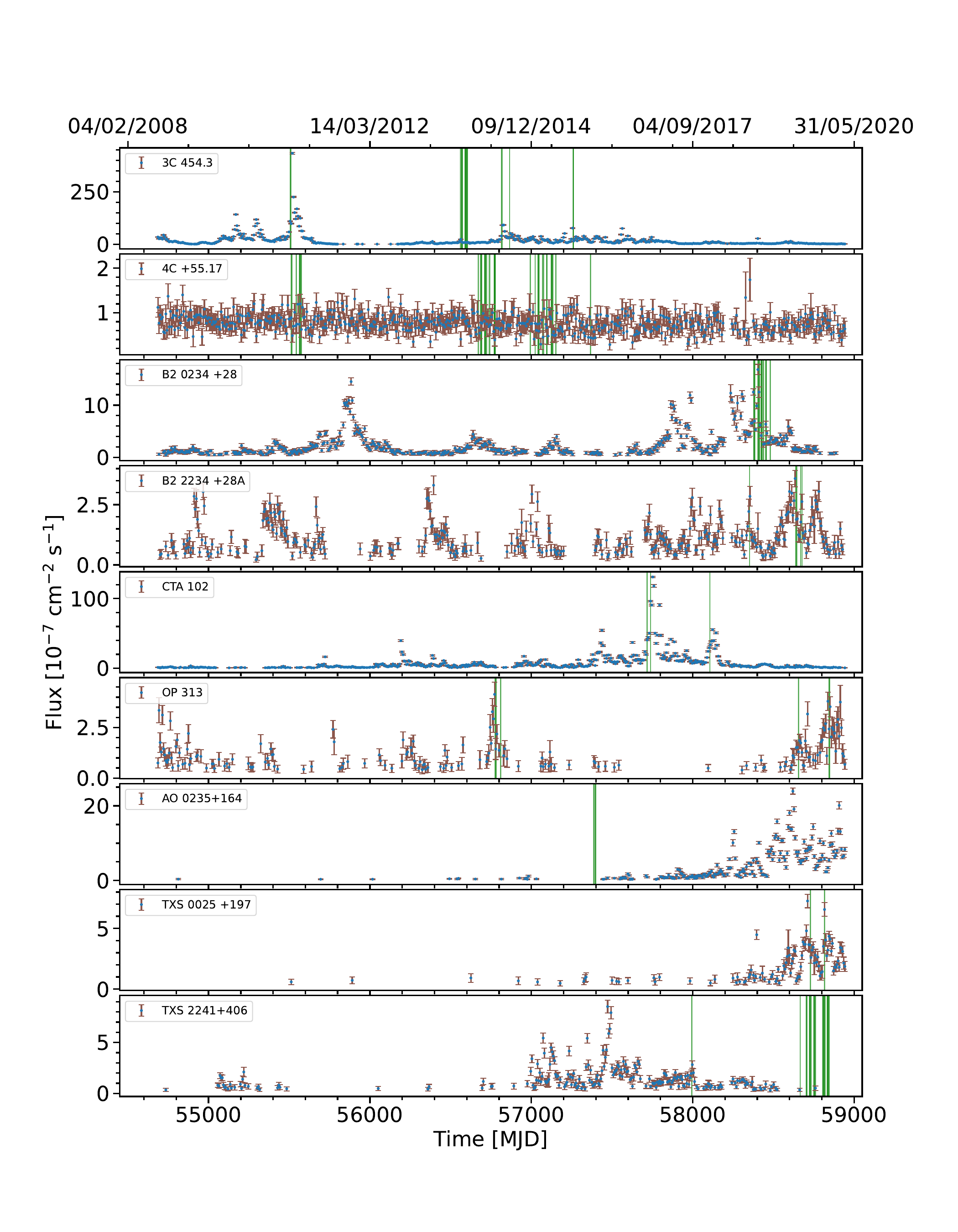}
\caption{$\textsl{Fermi}$-LAT light curve representing the flux from 2008 to 2020. For clarirty, only flux values (with TS>10) are shown in the plot, ULs are not reported. The green vertical areas on the graph indicate the specific days when MAGIC observations were carried out. This data are presented in weekly bins with an energy range of 0.1 - 1000 GeV.}
\label{fig:plot_LC_Fermi_12year}
\end{figure*}

\begin{figure*}
\includegraphics[scale=0.35]{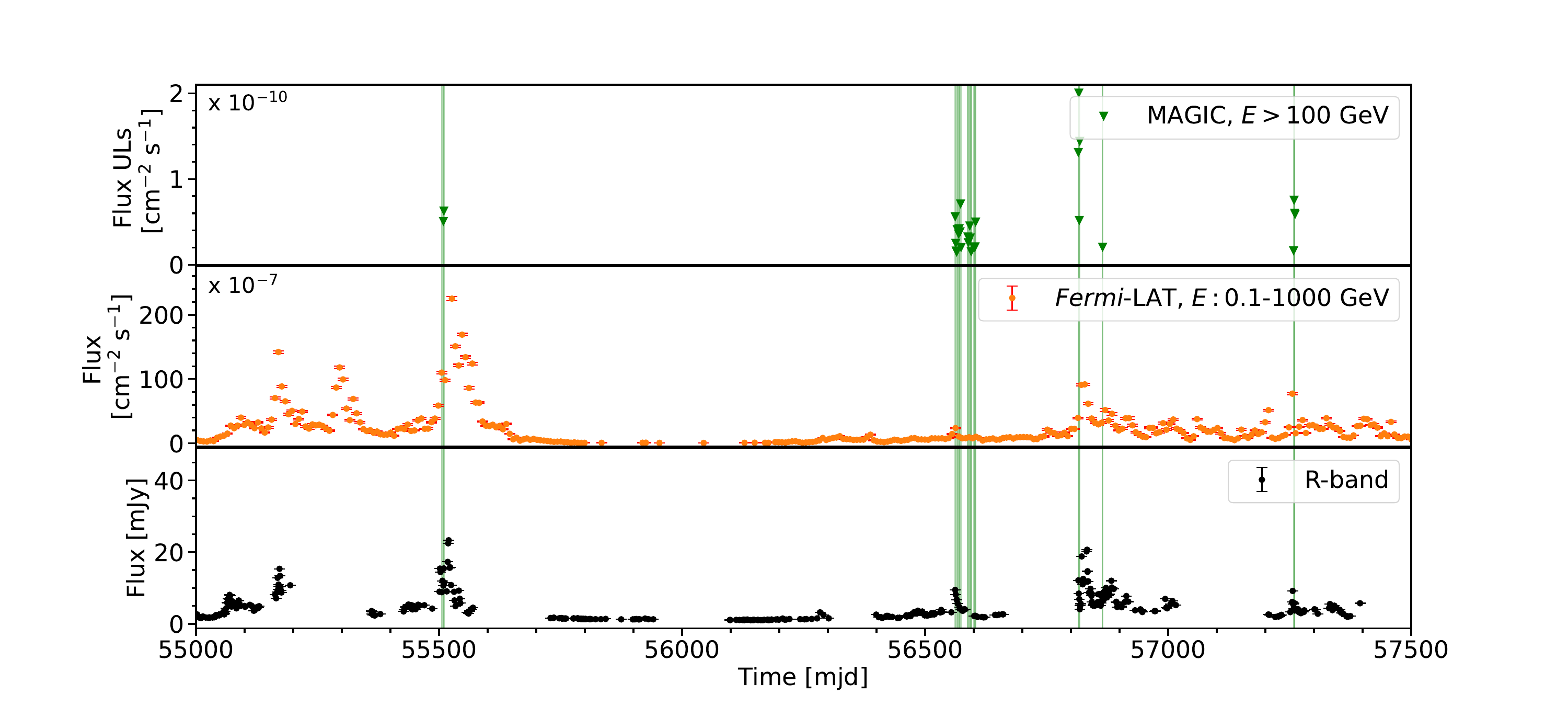}
\caption{MWL light curve of 3C 454.3. It shows the observations carried by MAGIC (top panel), $\textsl{Fermi}$-LAT (middle panel) and the R-band of \textit{Swift}-UVOT (bottom panel). The green vertical areas indicate the days during which MAGIC observations were carried out. $\textsl{Fermi}$-LAT light curve is presented in weekly bins and for the energy range of 0.1 - 1000 GeV. For clarirty, only $\textsl{Fermi}$-LAT flux values (with TS>10) are shown in the plot, ULs are not reported.}
\label{fig:plot_mwl_3C454}
\end{figure*}

\begin{figure*}
\includegraphics[scale=0.35]{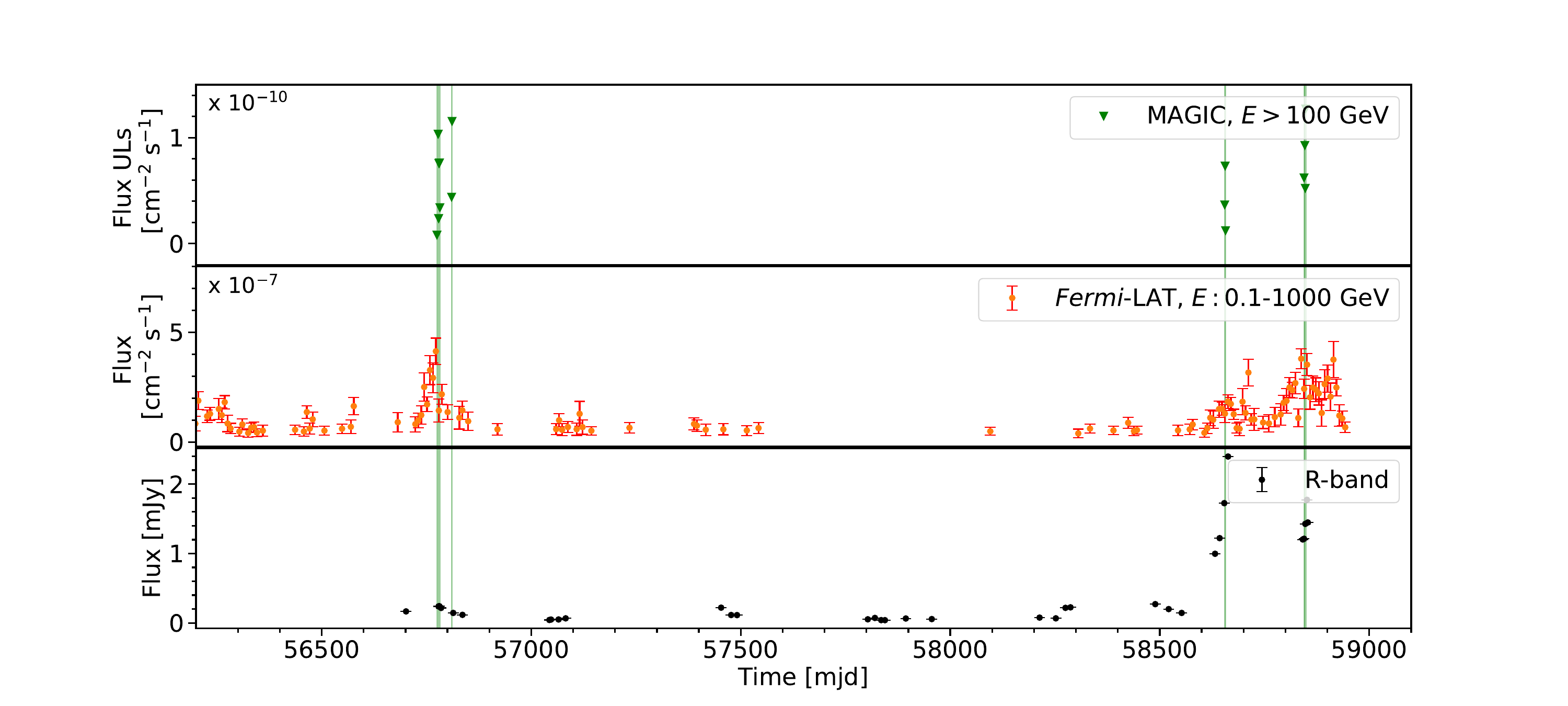}
\caption{MWL light curve of OP 313. The labels are the same as Fig. \ref{fig:plot_mwl_3C454}.}
\label{fig:plot_mwl_op313}
\end{figure*}

\begin{figure*}
\includegraphics[scale=0.35]{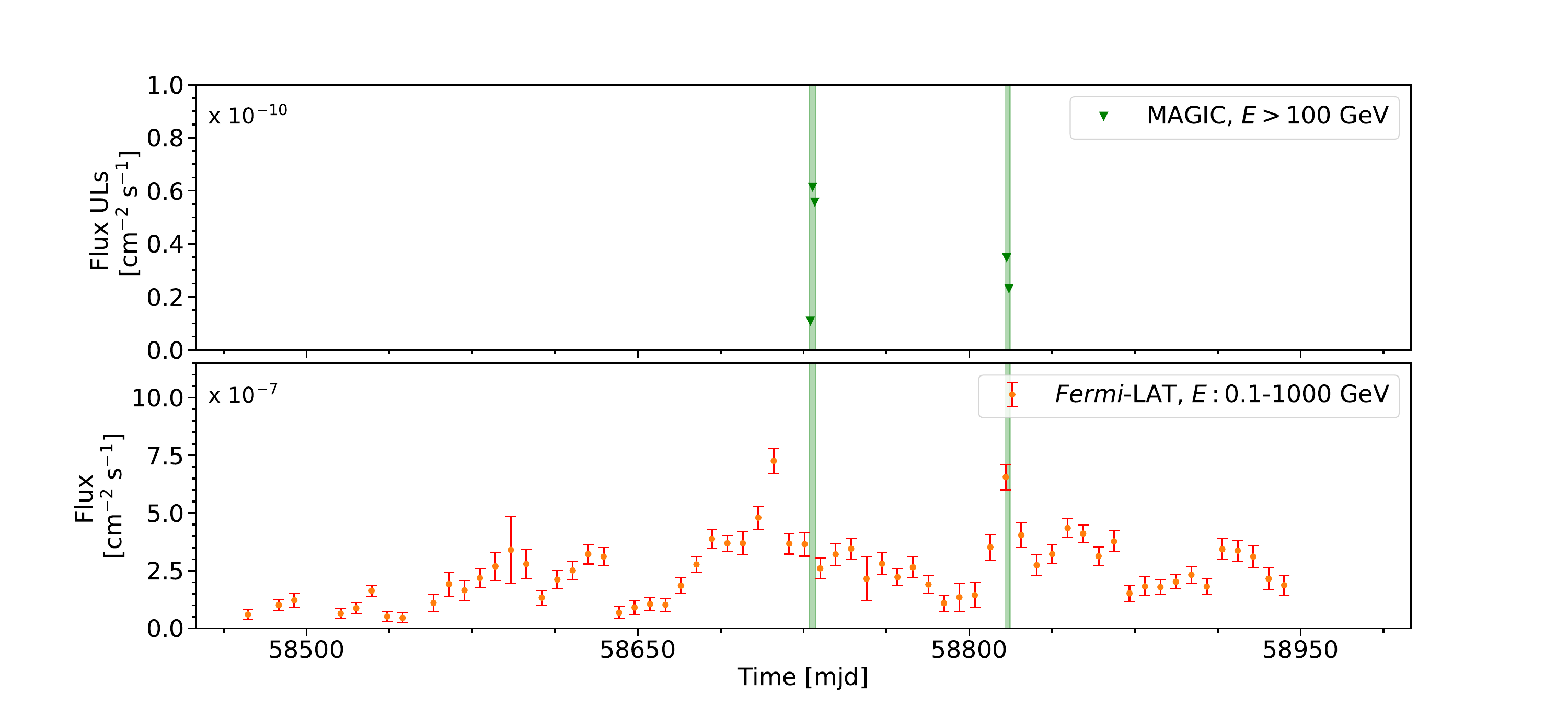}
\caption{MWL light curve of TXS 0025+197. The labels are the same as Fig. \ref{fig:plot_mwl_3C454}.}
\label{fig:plot_txs0025}
\end{figure*}

\begin{figure*}
\includegraphics[scale=0.35]{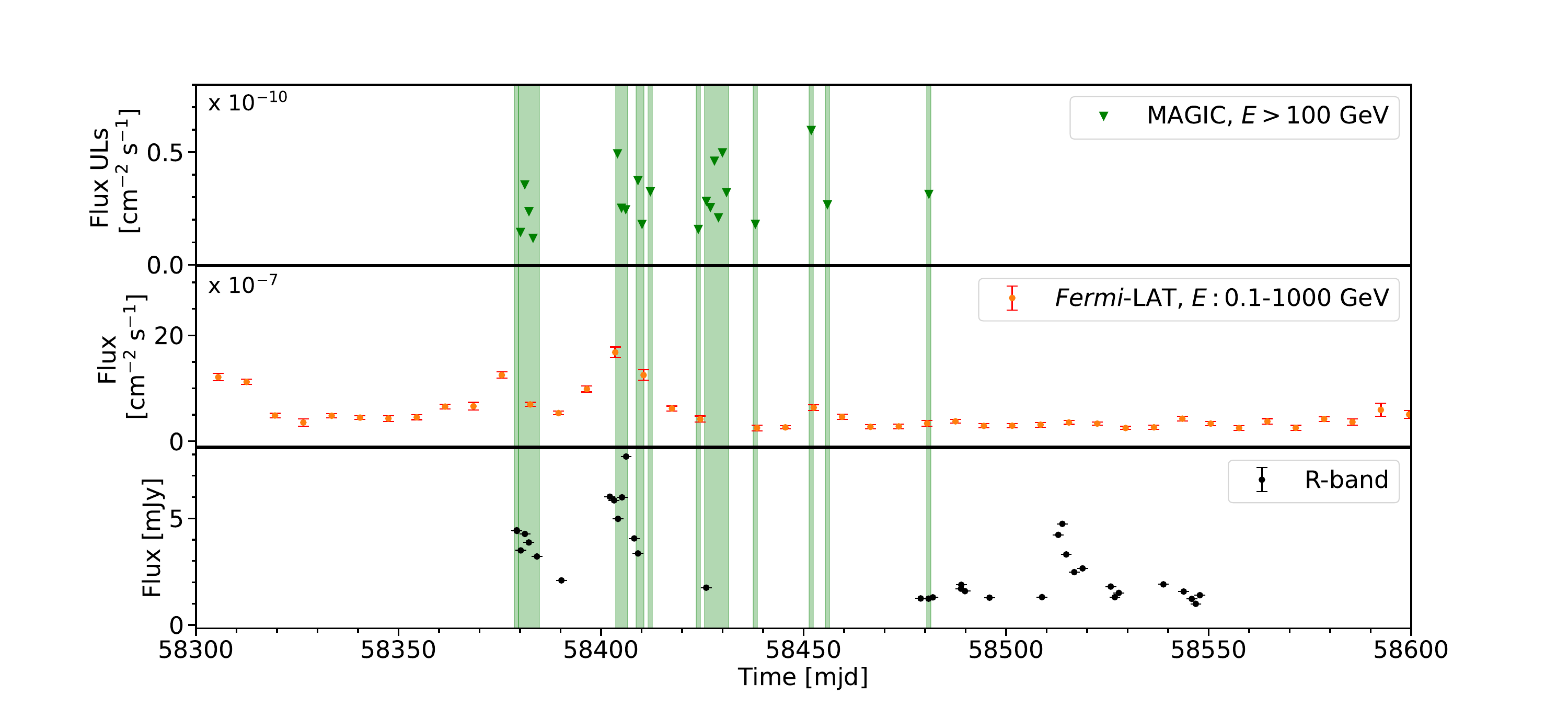}
\caption{MWL light curve of B2 0234+28. The labels are the same as Fig. \ref{fig:plot_mwl_3C454}.}
\label{fig:plot_mwl_b20234}
\end{figure*}

\begin{figure*}
\includegraphics[scale=0.35]{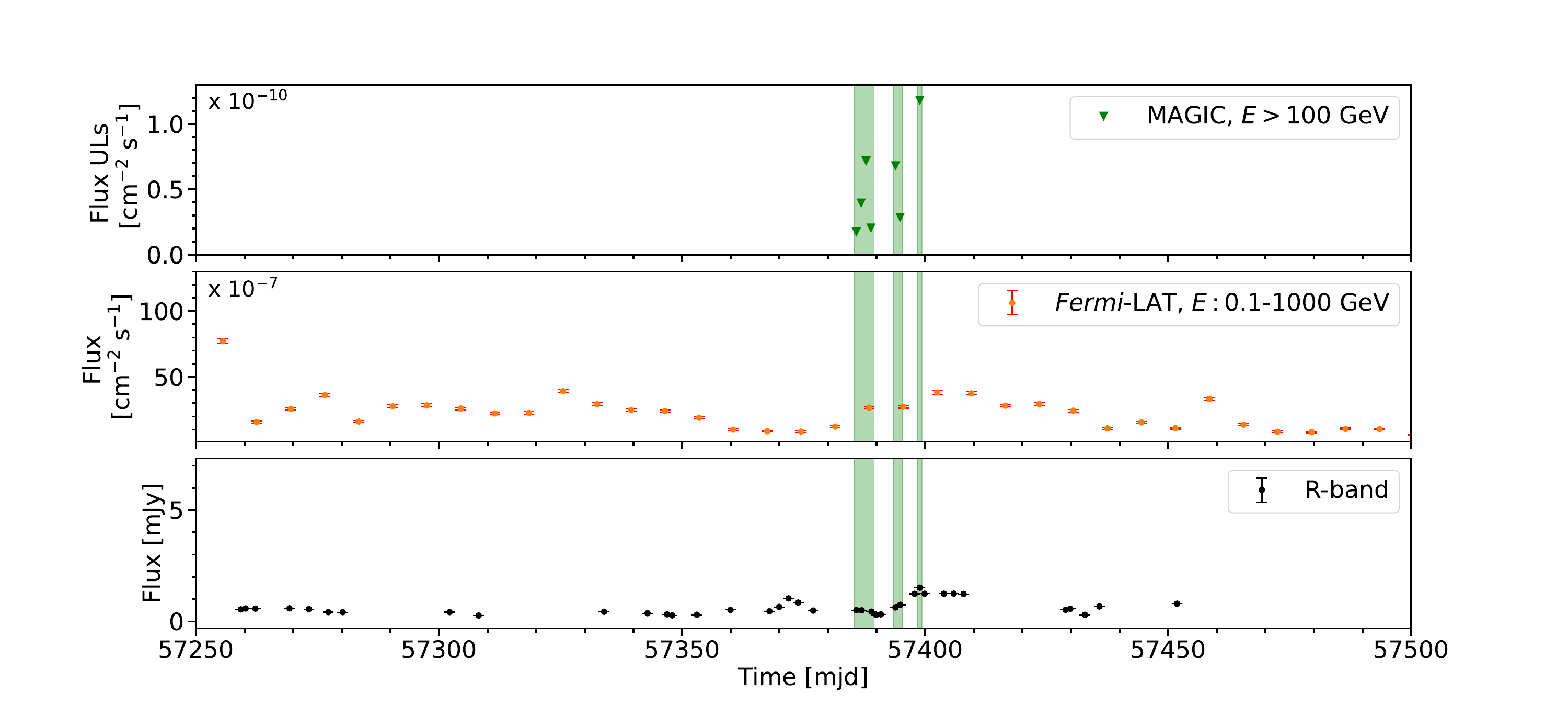}
\caption{MWL light curve of AO 0235+16. The labels are the same as Fig. \ref{fig:plot_mwl_3C454}.}
\label{fig:plot_mwl_ao}
\end{figure*}

\begin{figure*}
\includegraphics[scale=0.35]{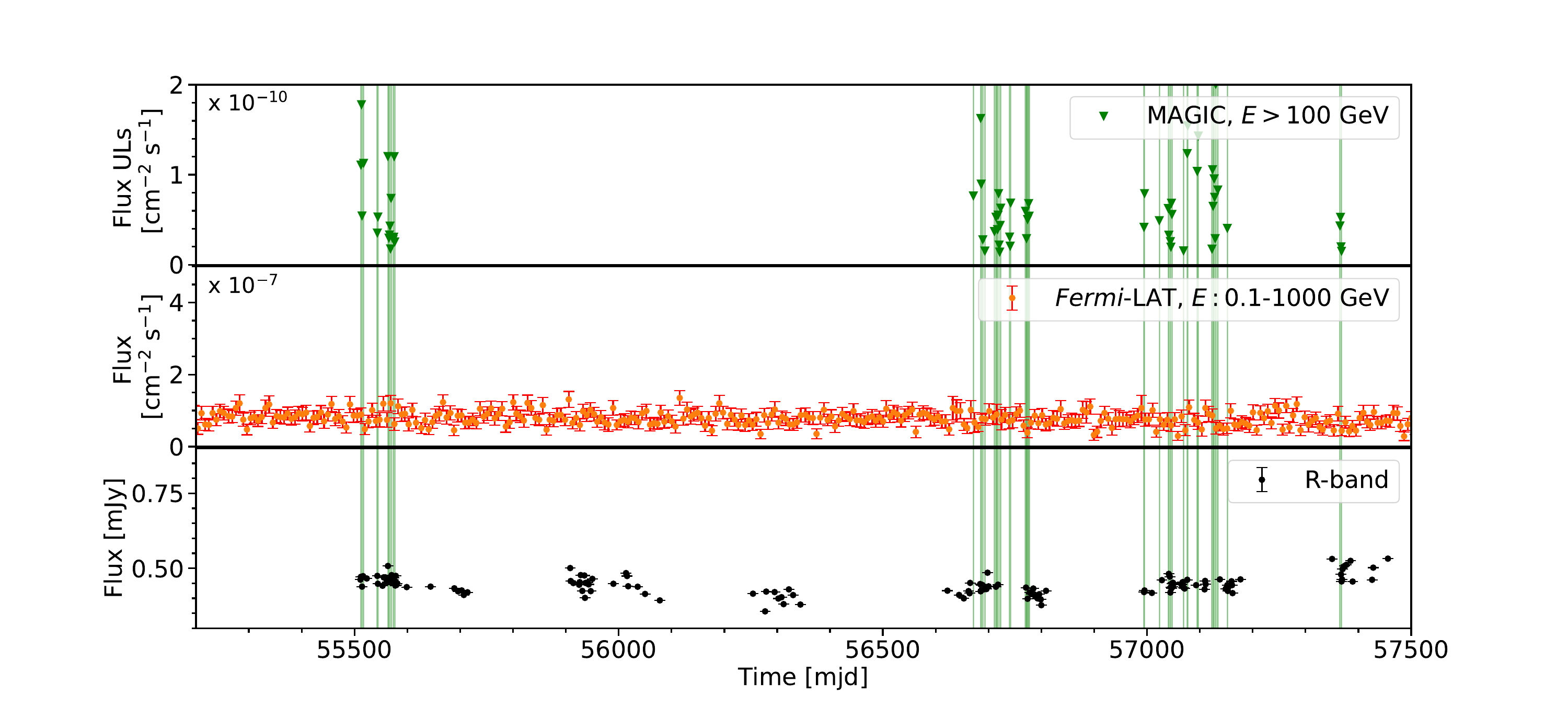}
\caption{MWL light curve of 4C 55.17. The labels are the same as Fig. \ref{fig:plot_mwl_3C454}.}
\label{fig:plot_mwl_4C55}
\end{figure*}

\begin{figure*}
\includegraphics[scale=0.35]{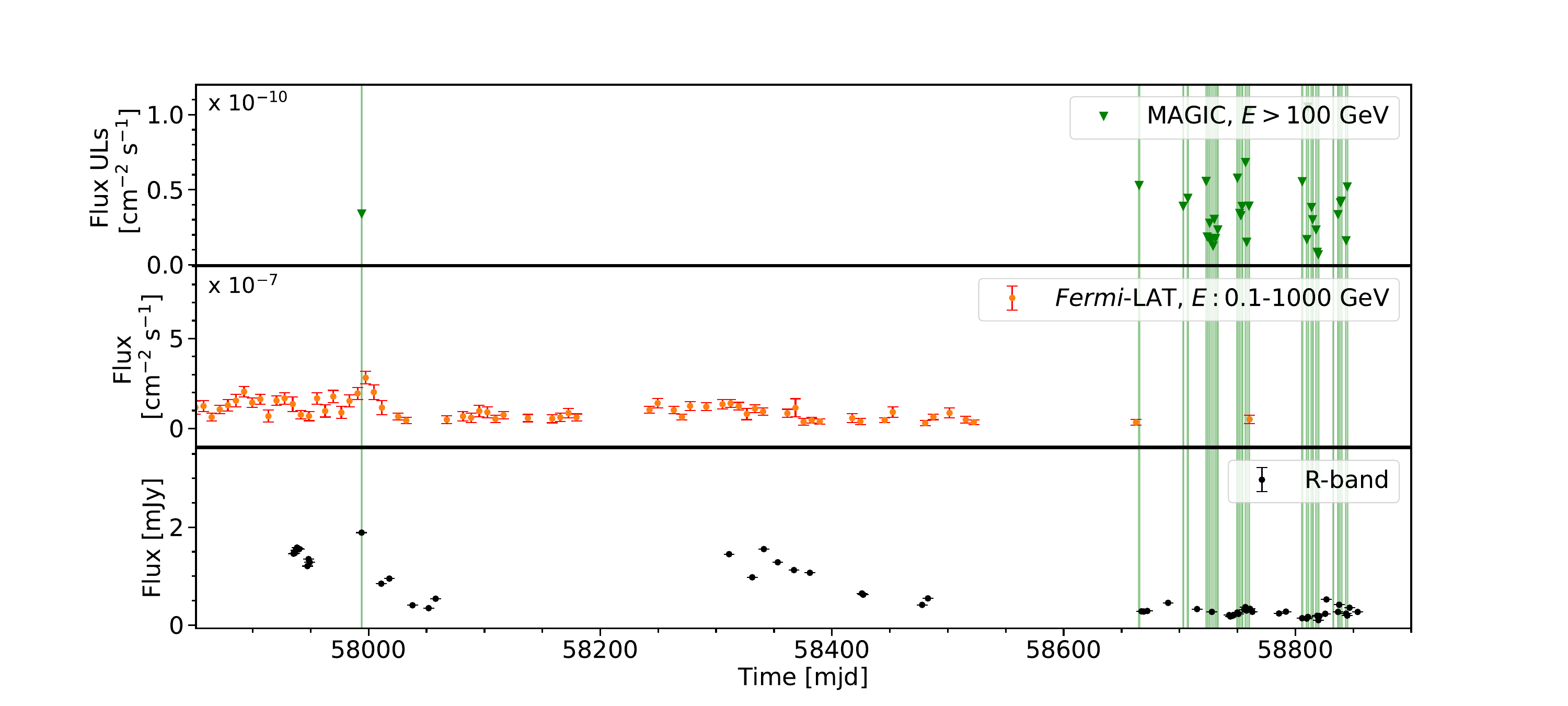}
\caption{MLW light curve of TXS 2241+406. The labels are the same as Fig. \ref{fig:plot_mwl_3C454}.}
\label{fig:plot_txs_2241}
\end{figure*}

%
%

\begin{table*}
	\centering
	\caption{Days when the MAGIC observations were carried out. These represent the time-interval used for the \textit{Fermi}-LAT analysis (see Sect. \ref{sec:fermi}). The one-day intervals are centered on the MAGIC observations. If observations occurred on consecutive days, then the integrated period, from the first day of observation to the last consecutive day of observation, is reported.}
	\label{tab:magic_fermi_obs}
	\begin{tabular}{l|p{11cm}} 
		\hline
		Association name  & MJDs - 50000 \\
		\hline
		TXS 0025+197 & 8727.56--8730.56, 8816.40--8818.40\\
		B2 0234+28 & 8378.60--8379.60, 8379.62--8384.62, 8403.56--8406.56, 8408.61--8410.61, 8411.68--8412.68, \newline 8423.51--8424.51, 8425.49--8431.49, 8437.59--8438.59, 8451.39--8452.39, 8455.39--8456.39, \newline 8480.43--8481.43\\
		AO 0235+16 & 7385.35--7389.35, 7393.42--7395.42, 7398.40--7399.40\\
		4C +55.17 & 5511.74--5514.74, 5516.73--5517.73, 5542.73--5544.73, 5562.68--5563.68, 5564.70--5569.70, \newline 5572.68--5576.68, 6670.78--6671.78, 6684.70--6686.70, 6688.69--6689.69, 6692.57--6693.57, \newline 6710.58--6711.58, 6713.59--6715.59, 6716.58--6723.58, 6739.47--6742.47, 6769.46--6770.46, \newline 6771.46--6772.46, 6773.44--6774.44, 6775.44--6777.44, 6993.75--6995.75, 7022.75--7023.75, \newline 7039.70--7041.70, 7043.71--7047.71, 7068.65--7069.65, 7075.65--7077.65, 7094.58--7095.58, \newline 7096.59--7097.58, 7122.49--7125.49, 7126.50--7130.50, 7133.45--7134.45, 7151.43--7152.43, \newline 7364.72--7368.72 \\
		OP 313 & 6774.51--6775.51, 6777.42--6782.42, 6809.47--6811.47, 8654.42--8657.42, 8843.76--8844.76, \newline 8845.72--8848.72 \\
		CTA 102 & 7715.40--7717.40, 7737.33--57739.33, 7748.34--57749.34, 8105.32--8106.32\\
		B2 2234+28A & 8351.66--8352.66, 8638.69--8639.69, 8642.68--8644.68, 8667.63--8668.63, 8676.67--8677.67\\
		TXS 2241+406 & 7993.56--7994.56, 8664.66--8665.66, 8702.69--8703.69, 8706.68--8707.68, 8722.48--8724.48, \newline 8725.53--8731.53, 8732.55--8733.55, 8749.51--8750.51, 8751.50--8754.50, 8756.49--8760.49, \newline 8805.43--8806.43, 8809.42--8811.42, 8813.41--8815.41, 8817.36--8820.36, 8832.35--8833.35, \newline 8836.32--8837.34, 8838.34--8840.34, 8843.33--8845.33\\
		3C 454.3 & 5505.47--5506.47, 5508.42--5510.42, 6561.41--6566.41, 6568.40--6570.40, 6571.38--6574.38, \newline
6587.37--6589.37, 6591.38--6593.38, 6594.36--6595.36, 6600.35--6601.35, 6602.35--6604.35, \newline
6814.67--6818.67, 6864.67--6865.67, 7257.60--7259.60, 7259.65--7260.65 \\
		\hline
	\end{tabular}
\end{table*}

\bsp	
\label{lastpage}
\end{document}